\documentclass[traditabstract]{aa}

\usepackage{graphicx}
\usepackage{txfonts}
\usepackage{natbib}
\usepackage{xcolor}
\usepackage{xspace}
\usepackage[colorlinks=true,linkcolor=blue,citecolor=blue]{hyperref}
\usepackage{etoolbox}

\makeatletter

\patchcmd{\NAT@citex}
  {\@citea\NAT@hyper@{%
     \NAT@nmfmt{\NAT@nm}%
     \hyper@natlinkbreak{\NAT@aysep\NAT@spacechar}{\@citeb\@extra@b@citeb}%
     \NAT@date}}
  {\@citea\NAT@nmfmt{\NAT@nm}%
   \NAT@aysep\NAT@spacechar\NAT@hyper@{\NAT@date}}{}{}

\patchcmd{\NAT@citex}
  {\@citea\NAT@hyper@{%
     \NAT@nmfmt{\NAT@nm}%
     \hyper@natlinkbreak{\NAT@spacechar\NAT@@open\if*#1*\else#1\NAT@spacechar\fi}%
       {\@citeb\@extra@b@citeb}%
     \NAT@date}}
  {\@citea\NAT@nmfmt{\NAT@nm}%
   \NAT@spacechar\NAT@@open\if*#1*\else#1\NAT@spacechar\fi\NAT@hyper@{\NAT@date}}
  {}{}

\makeatother

\newcommand{\hlink}[1]{\url{http://#1}\xspace}

\newcommand{\rfig}[1]{Fig.~\ref{#1}}
\newcommand{\rfigs}[1]{Figs.~\ref{#1}}
\newcommand{\req}[1]{Eq.~\ref{#1}}

\newcommand{\rtab}[1]{Table \ref{#1}}

\newcommand{\rapp}[1]{Appendix \ref{#1}}

\newcommand{\rsec}[1]{section \ref{#1}}
\newcommand{\rsecs}[1]{sections \ref{#1}}

\newcommand*\dd{\ensuremath{{\rm d}}}

\newcommand{\herschel}{{\it Herschel}\xspace}
\newcommand{\spitzer}{{\it Spitzer}\xspace}
\newcommand{\hubble}{{\it Hubble}\xspace}
\newcommand{\hst}{{\it HST}\xspace}

\newcommand{\um}{\mu{\rm m}}
\newcommand{\uJy}{\mu{\rm Jy}}

\newcommand{\sfr}{{\rm SFR}}
\newcommand{\sfruv}{{\rm SFR}_{\rm UV}}

\newcommand{\sfrms}{{\rm SFR}_{\rm MS}}
\newcommand{\ssfr}{{\rm sSFR}}
\newcommand{\sfe}{{\rm SFE}}
\newcommand{\lir}{L_{\rm IR}}

\newcommand{\leight}{L_8}
\newcommand{\ireight}{{\rm IR8}}

\newcommand{\fpah}{f_{\rm PAH}}
\newcommand{\lsun}{{\rm L}_\odot}
\newcommand{\msun}{{\rm M}_\odot}
\newcommand{\kelvin}{{\rm K}}
\newcommand{\Mpc}{{\rm Mpc}}

\newcommand{\Gyr}{{\rm Gyr}}

\newcommand{\yr}{{\rm yr}}
\newcommand{\dex}{{\rm dex}}
\newcommand{\mstar}{M_\ast}
\newcommand{\logoh}{12 + \log_{10}({\rm O/H})}
\newcommand{\mdust}{M_{\rm dust}}
\newcommand{\mgas}{M_{\rm gas}}
\newcommand{\fgas}{f_{\rm gas}}

\newcommand{\tdust}{T_{\rm dust}}

\newcommand{\uvj}{$UVJ$\xspace}
\newcommand{\bzk}{$BzK$\xspace}
\newcommand{\sersic}{S\'ersic\xspace}
\newcommand{\galfit}{GALFIT\xspace}
\newcommand{\gimtod}{GIM2D\xspace}
\newcommand{\bt}{B/T}

\newcommand{\mdisk}{M_{\rm disk}}
\newcommand{\mbulge}{M_{\rm bulge}}
\newcommand{\chisqr}{\chi^2}
\newcommand{\hone}{\ion{H}{i}}
\newcommand{\htwo}{{\rm H}_2}
\newcommand{\gdr}{\delta_{\rm GDR}}
\newcommand{\mhtwo}{M_{\htwo}}
\newcommand{\mhone}{M_{\hone}}
\newcommand{\mean}[1]{\left<#1\right>}

\newcommand{\sextractor}{{\sc SExtractor}\xspace}

\defcitealias{schreiber2015}{S15}

\begin{document}

\title{Observational evidence of a slow downfall of star formation \\ efficiency in massive galaxies during the last $10\,\Gyr$}
\titlerunning{A slow downfall of star formation efficiency in massive galaxies}

\author{C.~Schreiber\inst{1,2}
    \and D.~Elbaz\inst{1}
    \and M.~Pannella\inst{1,3,9}
    \and L.~Ciesla\inst{4,5,1}
    \and T.~Wang\inst{1,6}
    \and A.~Koekemoer\inst{7}
    \and M.~Rafelski\inst{8}
    \and E.~Daddi\inst{1}
}

\institute{
    Laboratoire AIM-Paris-Saclay, CEA/DSM/Irfu - CNRS - Universit\'e Paris Diderot, CEA-Saclay, pt courrier 131, F-91191 Gif-sur-Yvette, France \\
    \email{cschreib@strw.leidenuniv.nl}
    \and Leiden Observatory, Leiden University, NL-2300 RA Leiden, The Netherlands
    \and Institut d'Astrophysique de Paris, UMR 7095, CNRS, UPMC Univ. Paris 06, 98bis boulevard Arago, F-75014 Paris, France
    \and University of Crete, Department of Physics, 71003 Heraklion, Greece
    \and Institute for Astronomy, Astrophysics, Space Applications and Remote Sensing, National Observatory of Athens, GR-15236 Penteli, Greece
    \and School of Astronomy and Space Sciences, Nanjing University, Nanjing, 210093, China
    \and Space Telescope Science Institute, 3700 San Martin Drive, Baltimore, MD 21218, USA
    \and NASA Postdoctoral Program Fellow, Goddard Space Flight Center, Code 665, Greenbelt, MD 20771
    \and Max-Planck-Institut f\"ur Extraterrestrische Physik, Giessenbachstrasse 1, D-85748 Garching, Germany
}

\date{Received 17 August 2015; accepted 20 January 2016}

\abstract {
In this paper we study the causes of the reported mass-dependence of the slope of $\sfr$--$\mstar$ relation, the so-called ``Main Sequence'' of star-forming galaxies, and discuss its implication on the physical processes that shaped the star formation history of massive galaxies over cosmic time. We make use of the near-infrared high-resolution imaging from the {\it Hubble Space Telescope} in the CANDELS fields to perform a careful bulge-to-disk decomposition of distant galaxies and measure for the first time the slope of the $\sfr$--$\mdisk$ relation at $z=1$. We find that this relation follows very closely the shape of the nominal $\sfr$--$\mstar$ correlation, still with a pronounced flattening at the high-mass end. This is clearly excluding, at least at $z=1$, the secular growth of quiescent stellar bulges in star-forming galaxies as the main driver for the change of slope of the Main Sequence. Then, by stacking the \herschel data available in the CANDELS field, we estimate the gas mass ($\mgas=\mhone+\mhtwo$) and the star formation efficiency ($\sfe\equiv\sfr/\mgas$) at different positions on the $\sfr$--$\mstar$ relation. We find that the relatively low $\sfr$s observed in massive galaxies ($\mstar > 5\times10^{10}\,\msun$) are caused by a decreased star formation efficiency, by up to a factor of $3$ as compared to lower stellar mass galaxies, and not by a reduced gas content. The trend at the lowest masses is likely linked to the dominance of atomic over molecular gas. We argue that this stellar-mass-dependent $\sfe$ can explain the varying slope of the Main Sequence since $z=1.5$, hence over $70\%$ of the Hubble time. The drop of $\sfe$ occurs at lower masses in the local Universe ($\mstar > 2\times10^{10}\,\msun$) and is not present at $z=2$. Altogether this provides evidence for a slow downfall of the star formation efficiency in massive Main Sequence galaxies. The resulting loss of star formation is found to be rising starting from $z=2$ to reach a level comparable to the mass growth of the quiescent population by $z=1$. We finally discuss the possible physical origin of this phenomenon.
}


\maketitle

\section{Introduction}

The observation of a tight relation between the star formation rate ($\sfr$) and the stellar mass ($\mstar$) of galaxies, also called the ``Main Sequence'' of star-forming galaxies \citep{noeske2007}, at $z\simeq0$ \citep{brinchmann2004,elbaz2007}, $z\simeq1$ \citep{noeske2007,elbaz2007}, $z\simeq2$ \citep{daddi2007-a,pannella2009-a,rodighiero2011,whitaker2012-a} $z=3$--$4$ \citep{daddi2009,magdis2010-b,heinis2013,schreiber2015,pannella2015} and even up to $z=7$ \citep[e.g.,][]{stark2009,bouwens2012,stark2013,gonzalez2014,steinhardt2014,salmon2015} suggested a new paradigm for galaxy evolution. The tightness of this correlation is indeed not consistent with the frequent random bursts induced by processes like major mergers of gas-rich galaxies, and favors more stable, long-lasting episodes of star formation \citep{noeske2007}.

Most studies focusing on this Main Sequence have measured the slope (in logarithmic space) of this correlation, and many different values were reported. A thorough compilation was recently published in \cite{speagle2014}, summarizing most measurements obtained so far. In particular, we can distinguish three kinds of measurements. First, measured slopes close to unity \citep[e.g.,][]{elbaz2007,daddi2007-a,pannella2009-a,peng2010}. Second, slopes \emph{shallower} than unity, typically $0.8$, and as low as $0.6$ (e.g., \citealt{noeske2007,karim2011,rodighiero2011,bouwens2012,steinhardt2014,speagle2014,pannella2015}). And finally, more recently a third group of studies actually advocate a broken power-law shape, or continuously varying slopes, where low-mass galaxies are well fitted with a slope of unity, and high mass galaxies exhibit much shallower (if not flat) slopes \citep[e.g.,][]{whitaker2012-a,magnelli2014,whitaker2014,ilbert2015,schreiber2015,lee2015,gavazzi2015}. This latter, more refined description could actually explain the diversity of slope measurements that were obtained so far. Indeed, depending on the stellar mass range covered by the sample, which is usually limited, as well as the chosen redshift window, fitting a single power law will yield different best-fit slopes.

A tempting interpretation of this broken power law is that low mass galaxies evolve with a unique star formation efficiency, as shown by their universal specific $\sfr$ ($\ssfr\equiv\sfr/\mstar$) \citep[see, e.g., the discussions in][]{ilbert2015,lee2015}. Higher mass galaxies, on the other hand, depart from this universal relation and show a reduced star formation activity, probably gradually declining toward a quiescent state. This picture is somehow in contradiction with the idea that massive galaxies must \emph{quench} rapidly \citep[e.g.,][]{peng2010}, a process that often involves violent episodes in the lifetime of the galaxy, e.g., strong active galactic nucleus (AGN) feedback \citep{silk1998}. Instead, such a slow decline toward the red cloud could be more consistent with less abrupt processes like ``radio-mode'' AGN feedback \citep{croton2006,bower2006}, ``halo quenching'' \citep{gabor2012}, where the infalling gas is heated up and prevented from forming stars, or ``morphological quenching'' \citep{martig2009}, where the drop of star formation activity is caused by the presence of a massive and dense stellar bulge that increases the differential rotation within the disk and prevents gas from fragmenting.

Each of these mechanisms impacts directly the gas content of the galaxy, either by expelling the gas outside of the galaxy (thereby reducing the gas fraction) or by preventing cooling and fragmentation (thereby reducing the star formation efficiency). Testing these hypotheses implies measuring directly the gas content of galaxies, which formally requires costly spectroscopic campaigns to measure the molecular hydrogen mass through the carbon monoxide (CO) low-J emission lines, and atomic hydrogen (often assumed to be negligible at high redshift) through the $21\,{\rm cm}$ line. While this has been done extensively at $z=0$ \citep[e.g.,][]{walter2008,leroy2009,saintonge2011,boselli2014}, so far only small samples have been observed at $z\geq1$ \citep[e.g.,][]{daddi2008,dannerbauer2009,daddi2010,daddi2010-a,tacconi2010,tacconi2013} and these are limited to the most massive galaxies at every redshift. To circumvent this observational limitation, an alternative approach has been commonly used in the recent literature \cite[e.g.,][]{magdis2012,magnelli2012-a,santini2014,scoville2014,bethermin2015-a,genzel2015}, where the gas mass is inferred from the \emph{dust} mass of a galaxy, assuming for example that a fixed fraction of the metals (e.g., $\sim 30\%$, as discussed in \rsec{SEC:mgas}) condenses to form dust grains, and with the knowledge of the gas-phase metallicity \citep[see, e.g.,][]{franco1986}. Measuring dust masses and metallicities is still observationally challenging, however these are available for substantially larger samples. In particular, dust masses can be robustly measured using far-infrared and sub-millimeter photometry, either through individual measurements or stacking of large galaxy samples. At moderate redshifts ($z\leq1$), the \herschel space telescope probes rest-frame wavelength sufficiently large to accurately constrain the Raleigh-Jeans tail of the dust emission, and can therefore provide good estimations of the dust mass.

One important fact about dust-based gas mass estimates is that they include by construction the contribution of all phases of hydrogen gas, atomic and molecular. This means in particular that the star formation efficiency that is derived from such measurements probes the depletion of the entire gas reservoir of the galaxy, including the intermediate step of conversion from atomic to molecular hydrogen, and therefore provide a global point of view of the gas consumption. Since the pioneering work of \cite{kennicutt1998}, this has been the standard measure of the star formation efficiency.  It was shown later that the molecular gas is better correlated with the $\sfr$ than atomic hydrogen in local spirals \citep[e.g.,][]{wong2002,bigiel2008,bigiel2011}. While separating the two components in statistically large samples of distant galaxies to study how they relate to the $\sfr$ would bring valuable insight on star-formation, this is out of the scope of the present paper.

Recently, \cite{abramson2014} put forward another, possibly simpler explanation for the ``bending'' of the Main Sequence. They argue that, because of the presence of old stellar bulges within massive galaxies, the total stellar mass becomes a poor proxy for the mass of gas available\footnote{Regardless of the presence of a bulge, a similar conclusion can be drawn from the absence of a strong correlation between surface densities of stars and gas in nearby galaxies; e.g., \cite{shi2011}.}. One should rather expect the star formation rate to correlate with the mass of the \emph{disk} instead, since this is where the star-forming gas is located. To support their claim, they used bulge-to-disk decompositions of the observed light profiles of local galaxies in the Sloan Digital Sky Survey (SDSS), and estimated their disk masses. They found indeed that the slope of the Main Sequence was put back to unity at all masses (at least for $\mstar > 10^{10}\,\msun$) if the disk mass was substituted to the total stellar mass (see, however, \citealt{guo2015} where a conflicting result is obtained using the same data set). In \cite{schreiber2015} (hereafter \citetalias{schreiber2015}), we have reported that the high-mass slope of the Main Sequence is gradually decreasing with time, departing from unity at $z<2$ and reaching the shallowest values in the present day \citep[see also][]{whitaker2014,lee2015,gavazzi2015}, which seems consistent with the progressive growth of bulges (see also \citealt{wuyts2011}, \citealt{whitaker2015} and \citealt{tacchella2015-a}).

Thanks to the very high angular resolution provided by the \hubble Advanced Camera for Surveys (ACS) imaging, it is possible to perform the morphological analysis of the stellar profile of distant galaxies out to $z=1$, either through non-parametric approaches \citep[e.g.,][]{abraham1996,conselice2003,ferguson2004,lotz2004}, profile fitting \citep[e.g.,][]{bell2004,ravindranath2004,barden2005,mcintosh2005,pannella2006,haeussler2007,pannella2009-a}, or decomposition of this profile into multiple components \citep[e.g.,][]{simard1999,simard2002,stockton2008}. The advent of the WFC3 camera on board \hubble has recently allowed studying the rest-frame near-IR (NIR) and optical stellar profiles toward higher redshifts \citep[e.g.,][]{vanderwel2012,newman2012,bruce2012,bruce2014,lang2014}. In particular, \cite{bruce2012} have performed bulge-to-disk decomposition on the CANDELS $H$-band imaging in the UDS field, focusing of massive galaxies ($\mstar > 10^{11}\,\msun$) from $z=1$ to $z=3$, and finding a clear trend of decreasing bulge-to-total ratio ($\bt$) with redshift. However, later on \cite{lang2014} pushed the analysis down by one order of magnitude in stellar mass in all five CANDELS fields. By fitting stellar-mass maps estimated through resolved SED-fitting, they derived the relation between $\mstar$ and $\bt$ for star-forming and quiescent galaxies, and found very little evolution of this relation with redshift. Both these observations are contradictory, and would potentially lead to different conclusions when trying to link the bulge mass to the Main Sequence bending.

Our goal in this paper is therefore to investigate directly the possible causes for the evolution of the slope. To do so, we analyze a sample of $z=1$ galaxies and follow two complementary approaches. On the one hand, we estimate the mass of the disks in each galaxy, and see if the $\sfr$--$\mdisk$ relation is linear, as found in the Local Universe. On the other hand, we estimate the gas masses in our sample and quantify the mass evolution of both the gas fraction ($\fgas$) and the star formation efficiency ($\sfe$) to see which of these two quantity best correlates with the bending of the Main Sequence.

Both studies are based on a common sample of $z=1$ galaxies drawn from the CANDELS fields \citep{grogin2011,koekemoer2011}, and we also use data from the Local Universe (the \herschel Reference Survey) to extend and confirm our results regarding the gas mass measurements. The precise sub-samples used in each study are detailed in \rsecs{SEC:sample:gz1} and \ref{SEC:sample:gz0} for the gas mass study, and \rsec{SEC:sample:mz1} for the disk mass study. In \rsec{SEC:mdisk_base} we describe how we perform the bulge-to-disk decomposition to measure the stellar mass of the disk, while in \rsec{SEC:sfe} we describe the procedure we employ to measure the gas masses. Our results are then presented in \rsec{SEC:results}.

In the following, we assume a $\Lambda$CDM cosmology with $H_0 = 70\ {\rm km}\,{\rm s}^{-1} {\rm Mpc}^{-1}$, $\Omega_{\rm M} = 0.3$, $\Omega_\Lambda = 0.7$ and, unless otherwise specified, a \cite{salpeter1955} initial mass function (IMF) to derive both star formation rates and stellar masses. All magnitudes are quoted in the AB system, such that $M_{\rm AB} = 23.9 - 2.5\log_{10}(S_{\!\nu}\ [\uJy])$. Finally, the gas masses that we derive include the contribution of helium.

\section{Samples and galaxy properties}

\begin{figure}
    \centering
    \includegraphics*[width=9cm]{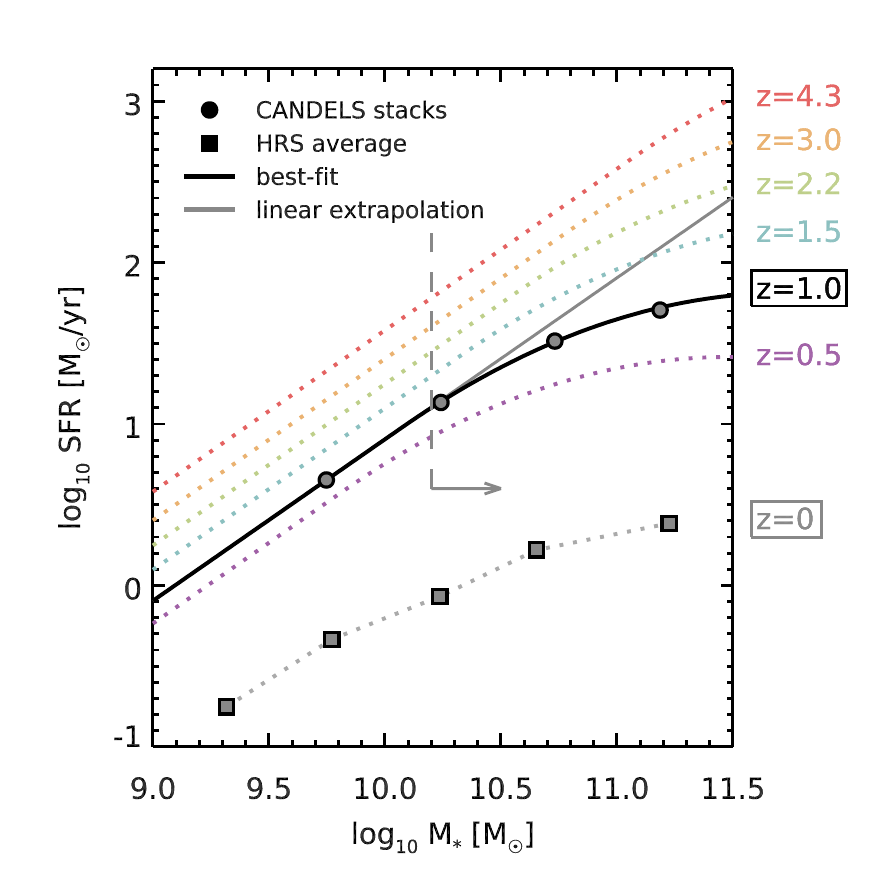}
    \caption{The Main Sequence of star-forming galaxies at different redshifts. Solid circles and fits (solid black line and dotted colored lines) are taken from \citetalias{schreiber2015}. Statistical error bars are smaller than the symbols. In the present work, we focus on a redshift range around $z=1$, which is highlighted in this plot. There, to illustrate the change of slope of the Main Sequence, we show with a gray solid line the extrapolation of the low-mass $\ssfr\equiv\sfr/\mstar$, with a slope of unity. The gray dashed line and the arrow indicate the region of this diagram within which we perform the morphological decomposition of the \hst light profiles of $z=1$ galaxies (\rsec{SEC:sample:mz1}). We also show for reference the Main Sequence as seen in the HRS at $z=0$ \citep[see][]{ciesla2016}.}
    \label{FIG:sfrms_sample}
\end{figure}

\begin{table*}
\caption{Summary of the various samples used in this paper. \label{TAB:sample}}
\begin{center}
\begin{tabular}{lccccc}
\hline \\[-0.35cm]
Sample                 & Number\tablefootmark{a}  & $\mstar$ & \uvj\tablefootmark{b} & IR\tablefootmark{c}       & Robust $\bt$\tablefootmark{d}    \\
                                      &           & $\msun$            &          &          &     \\
\hline\hline \\[-0.35cm]
{\bf Morphological decomposition ($z=1$)} \\
$H$-sample             & $2\,439$  & $> 2\times10^{10}$ & no       & no       & $2\,081$ ($85$\%)  \\
UVJ-SF                 & $1\,499$  & $> 2\times10^{10}$ & yes      & no       & $1\,280$ ($85$\%)  \\
IR-sample              & $946$     & $> 2\times10^{10}$ & yes      & yes      & $783$    ($83$\%)  \\
IR-sample + good $\bt$ & $783$     & $> 2\times10^{10}$ & yes      & yes      & $100$\% \\
{\bf Gas mass measurement} \\
CANDELS ($z=1$)        & $4\,730$  & $> 3\times10^{9}$  & yes      & no       & ... \\
HRS ($z=0$)            & $131$     & $> 10^{9}$         & yes      & no       & ... \\
\hline
\end{tabular}
\end{center}
\tablefoot{We distinguish two sets of samples. First, we list the $z=1$ samples we use to study the bulge-to-disk decompositions (\rsec{SEC:mdisk_base}). Each step of the selection process corresponds to a different row; the corresponding stellar mass distributions are shown in \rfig{FIG:mstar_hist}. Second, we show the two $z=1$ and $z=0$ samples involved in the gas content measurements (\rsec{SEC:sfe}). \\
\tablefoottext{a}{Number of galaxies in the sample} \tablefoottext{b}{Indicates if the sample is \uvj-selected.} \tablefoottext{c}{Indicates if the sample is IR-selected, i.e., contains only galaxies individually detected by \spitzer MIPS and/or \herschel.} \tablefoottext{d}{Fraction of the galaxies in the sample with a robust bulge-to-disk decomposition.}}
\end{table*}

In this work we investigate the change of slope of the Main Sequence from two different angles. Both approaches require different samples that, even if drawn from the same data set, differ noticeably in terms of their stellar mass and star formation rate completeness. For this reason these samples and their corresponding selections are summarized in \rtab{TAB:sample}.

On the one hand, we measure the gas content inside Main Sequence galaxies to look for a decrease of either the gas fraction or the star formation efficiency. To do so, we use the stacked \herschel SEDs of \citetalias{schreiber2015} at $z=1$ in the CANDELS fields (see \rsec{SEC:sample:gz1}) to measure both the SFR and the gas masses. This sample contains all star-forming galaxies at $0.7<z<1.3$ with $\mstar \ge 3\times10^9\,\msun$, and is complete both in stellar mass and $\sfr$ above this threshold. We complement this analysis with a $z=0$ sample of Main Sequence galaxies from the HRS (see \rsec{SEC:sample:gz0}), which is volume-limited.

On the other hand, we extract a subsample of massive galaxies ($\mstar \ge 2\times10^{10}\,\msun$) from our $z\sim1$ sample and perform the morphological decomposition of their \hst light profile. Among these, we mostly consider the galaxies with an \emph{individual} IR detection in order to derive robust $\sfr$s for each object, yielding a subsample that is both mass- and $\sfr$-selected. The description of this subsample is given in \rsec{SEC:sample:mz1}.

For a description of the fields and the photometry, as well as the method used to measure physical properties such as redshifts, stellar masses and star formation rates, we refer the reader to the papers where these samples were initially introduced (i.e., \citetalias{schreiber2015} and \citealt{ciesla2016}).

\subsection{CANDELS sample for the gas mass measurements at $z=1$ \label{SEC:sample:gz1}}

For the gas mass measurements at $z=1$, we use the stacked \herschel photometry in the CANDELS fields presented in \citetalias{schreiber2015}. In this work, we showed that the bending of the Main Sequence is more pronounced at lower redshifts, and almost absent by $z>2$ (see also \rfig{FIG:sfrms_sample}). To study the origin of this bending, we therefore need to focus on low redshifts, where the bending is most significant. On the other hand, the area covered by the CANDELS fields is relatively small, and consequently we cannot afford to reach too low redshifts, say $z<0.5$, without being affected by limited statistics and small volumes. Furthermore, our estimation of the gas mass is based on the dust mass (see \rsec{SEC:mgas}), and at $z>1.5$ \herschel does not probe the Rayleigh-Jeans tail of the dust SED ($\lambda_{\rm rest} > 250\,\um$), which would prevent accurate determination of the dust mass \citep{scoville2014}.

For these reasons we choose to base our analysis on galaxies at $0.7 < z < 1.3$, and use the same sample as in \citetalias{schreiber2015}, namely selecting all the galaxies in this redshift window that are classified as \uvj star-forming:
\begin{equation}
    UVJ_{\rm SF} = \left\{\begin{array}{rcl}
        U - V &<& 1.3\,\text{, or} \\
        V - J &>& 1.6\,\text{, or} \\
        U - V &<& 0.88\times(V - J) + 0.49\,.
    \end{array}\right.\label{EQ:uvj}
\end{equation}
This selection is illustrated later in \rfig{FIG:uvj_bt}. As discussed in \citetalias{schreiber2015}, more than $85\%$ of the \herschel detections are classified as \uvj star-forming. The \uvj selection is therefore an efficient tool to pinpoint star-forming galaxies, even when MIR or FIR detections are lacking. However, it affects more strongly the galaxies at high stellar mass. In particular, between $10^{11}$ and $3\times10^{11}\,\msun$, about half of our galaxies are classified as \uvj quiescent. Since the precise definition of \req{EQ:uvj} could affect our results, we discuss its impact {\it a posteriori} in \rapp{SEC:uvj}.

\subsection{HRS sample for the gas mass measurements in the Local Universe \label{SEC:sample:gz0}}

For the $z=0$ sample, we define the dividing line between ``star-forming'' and ``quiescent'' galaxies as follows:
\begin{equation}
    UVJ_{\rm SF}\text{ (HRS)} = \left\{\begin{array}{rcl}
        U - V &<& 1.6\,\text{, or} \\
        V - J &>& 1.6\,\text{, or} \\
        U - V &<& 0.88\times(V - J) + 0.79\,.
    \end{array}\right.\label{EQ:uvj_hrs}
\end{equation}
In practice, this is equivalent to making a cut in $\ssfr > 6\times10^{-3}\,\Gyr^{-1}$, i.e., about one dex below the $z=0$ Main Sequence. Different \uvj dividing lines have been adopted in the literature, reflecting a combination of both zero point offsets in the photometry and physical evolution of the colors caused by the evolution of the $\ssfr$. For example, \cite{williams2009} used different \uvj classifications depending on the redshift, with a $0<z<0.5$ criterion that is different from \req{EQ:uvj_hrs} by only $0.1$ magnitudes, and a $1<z<2$ criterion identical to our \req{EQ:uvj}.

In the following, we use all the galaxies from the HRS survey that satisfy the \uvj criterion given above, regardless of their morphological type. In practice, the \uvj selection naturally filters out all the early-type galaxies (E-S0-S0/Sa), and about half of the \hone-deficient galaxies (as defined in \citealt{boselli2010}).

However, it is important to note that, although the HRS is a purely $K$-band selected sample, the volume it spans is relatively small and the HRS is thus subject to cosmic variance. Furthermore, because one of the science goals of the HRS is to study the influence of the environment on the star formation activity, the sample also contains the Virgo cluster, a strong overdensity that encloses $46\%$ of the galaxies in the whole HRS (and $39\%$ of \uvj star-forming galaxies). This is a very biased environment, and although clusters are more common in the Local Universe, the HRS is known to be particularly deficient in gas mass, likely because of the inclusion of Virgo \citep{boselli2010}. To ease the comparison with our $z=1$ sample described in the previous section, we therefore exclude from the HRS all the galaxies that belong to Virgo ($149$ galaxies out of $323$). Combined with the \uvj selection, this excludes $80\%$ of the \hone-deficient galaxies, and yields a final sample of $131$ galaxies. We note however that our results would be essentially unchanged if we were to keep the Virgo galaxies in our sample.

\subsection{CANDELS sample for the morphological decompositions at $z=1$ \label{SEC:sample:mz1}}

\begin{figure}
    \centering
    \includegraphics*[width=9cm]{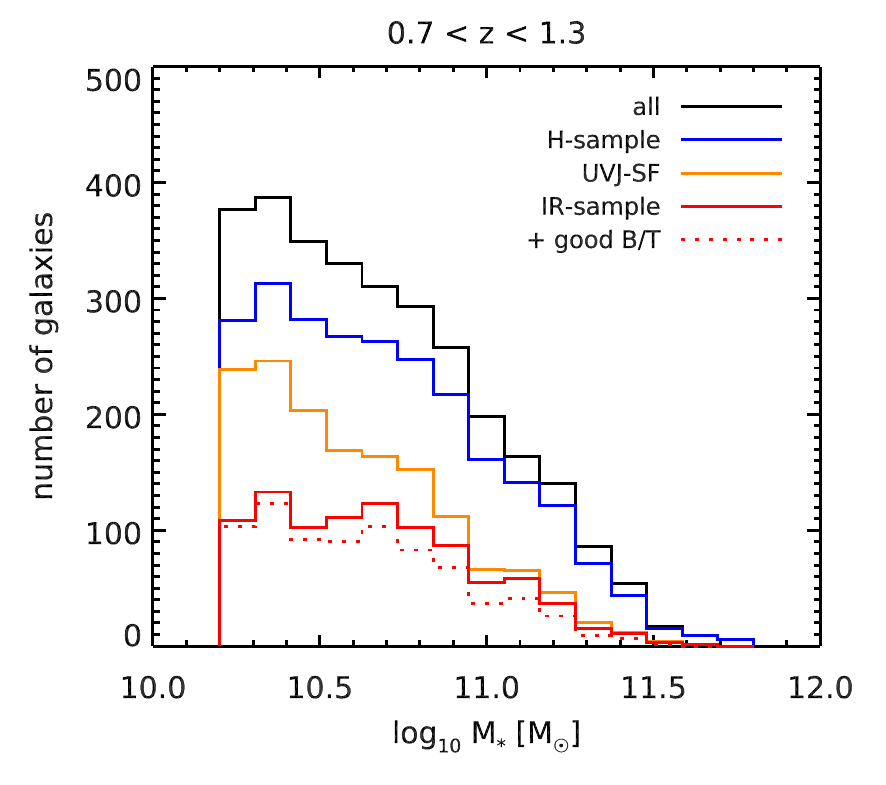}
    \caption{Stellar mass distribution of the various samples at $z=1$ that we consider for the morphological decomposition (\rsec{SEC:sample:mz1}). Each line corresponds to a step of the selection process, progressively decreasing the number of objects in the sample as in \rtab{TAB:sample}. The black solid line shows the distribution of our parent sample, containing all the galaxies at $0.7 < z < 1.3$ with $\mstar > 2\times10^{10}\,\msun$ and accurate determination of both redshift and stellar mass. The blue solid line is our $H$-sample, after removing close pairs and IRAC power-law AGNs from the parent sample. The orange solid line shows galaxies in the $H$-sample that are classified as \uvj star-forming (\req{EQ:uvj}). The red solid line is our IR-sample of galaxies with a MIR or FIR detection. Finally, the dotted line indicates the number of those galaxies for which we can reliably decompose the light profile.}
    \label{FIG:mstar_hist}
\end{figure}

For the morphological analysis, we consider the same redshift window as for the gas mass measurement at $z=1$, following the same motivations. In addition, limiting ourselves to $z=1$ ensures that the \hst $H$ band probes the rest-frame $i$ band, where mass-to-light ratios are weakly varying \citep[e.g.,][]{dejong1996}. However, to obtain reliable morphological decompositions, we need to select galaxies that are sufficiently bright and without strong contamination from neighboring objects. The various steps of the selection described below are illustrated on the stellar mass distribution in \rfig{FIG:mstar_hist}.

We thus select galaxies more massive than $2\times10^{10}\,\msun$, corresponding roughly to an $H$-band limited sample at these redshifts, with no galaxy fainter than $H = 22.5$ (see \rapp{APP:simu} where we justify this choice using simulated images). Unfortunately, this stellar mass cut will prevent us from performing the morphological decomposition in the regime where the Main Sequence is linear, as shown in \rfig{FIG:sfrms_sample}. However, it is known that disk-dominated galaxies dominate the low-mass galaxy population, both in the Local Universe \citep[e.g.,][]{bell2003} and at higher redshifts \citep[e.g.,][]{pannella2009-a,lang2014,bluck2014-a}. Therefore we will assume in the following that galaxies at $\mstar < 2\times10^{10}\,\msun$ are disk-dominated, with $\mstar \simeq \mdisk$, and only consider changes in Main Sequence slope above this threshold. We also remove $6$ IRAC power law AGNs \citep[following][]{donley2012}.

To prevent systematic effects in the morphological analysis due to strong galaxy blending (either due to mergers or chance projections), we also need to remove from our sample the galaxies that have too close bright neighbors in the $H$-band image. Therefore, we flagged the galaxies that have at least one companion within $2\arcsec$ with a total flux that is no less than $10\%$ fainter. This flags out $410$ galaxies, and our final ``$H$-sample'' consists of $2\,439$ galaxies ($1\,499$ of which are \uvj star-forming according to \req{EQ:uvj}).

Then, among these, we also consider the ``IR-sample'' that consists of star-forming galaxies with a MIR or FIR detection ($>$$5\,\sigma$), i.e., with a robust $\sfr$ estimate coming from \spitzer or \herschel observations. To do so, we first select star-forming galaxies using the \uvj diagram and \req{EQ:uvj}. Then, to derive the $\sfr$s, we start from the same IR catalogs as those introduced in \citetalias{schreiber2015}, but here we further revisit the catalogs to solve an issue that can have important consequences for the present study. Briefly, we flag the \spitzer MIPS detections that are potentially wrongly associated to their $H$-band counterparts because of the adopted source extraction procedure. The details of this flagging procedure are described in \rapp{APP:24clean}. In total we flag no more than $5\%$ of the MIPS detections in the catalog as wrong or uncertain associations. Two thirds of these are \uvj quiescent galaxies, and are therefore not part of the IR-sample.

The final IR-sample contains $947$ galaxies, and therefore $63\%$ of the star-forming galaxies of the $H$-sample have a robust $\sfr$ estimation (see \rfig{FIG:mstar_hist}). For consistency checks, we do perform the morphological detection on the whole $H$-sample (i.e., including in particular those galaxies that are \uvj quiescent), but only use the IR-sample to derive the slope of the Main Sequence, meaning that we will eventually work with a sample that is both mass- \emph{and} $\sfr$-selected. This is not an issue for our purposes. Even though half of the star-forming galaxies close to our stellar mass threshold are not seen in the MIR or FIR, the IR-sample is at least $80\%$ complete for star-forming galaxies above $\mstar > 5\times10^{10}\,\msun$ (see \rfig{FIG:mstar_hist}). Since the change of slope of the Main Sequence is most pronounced at the massive end, we will be able to witness any modification of this slope once the disk mass is substituted to the total stellar mass.

\section{Measuring disk masses in distant galaxies \label{SEC:mdisk_base}}

\begin{figure*}
    \centering
    \includegraphics*[width=18cm]{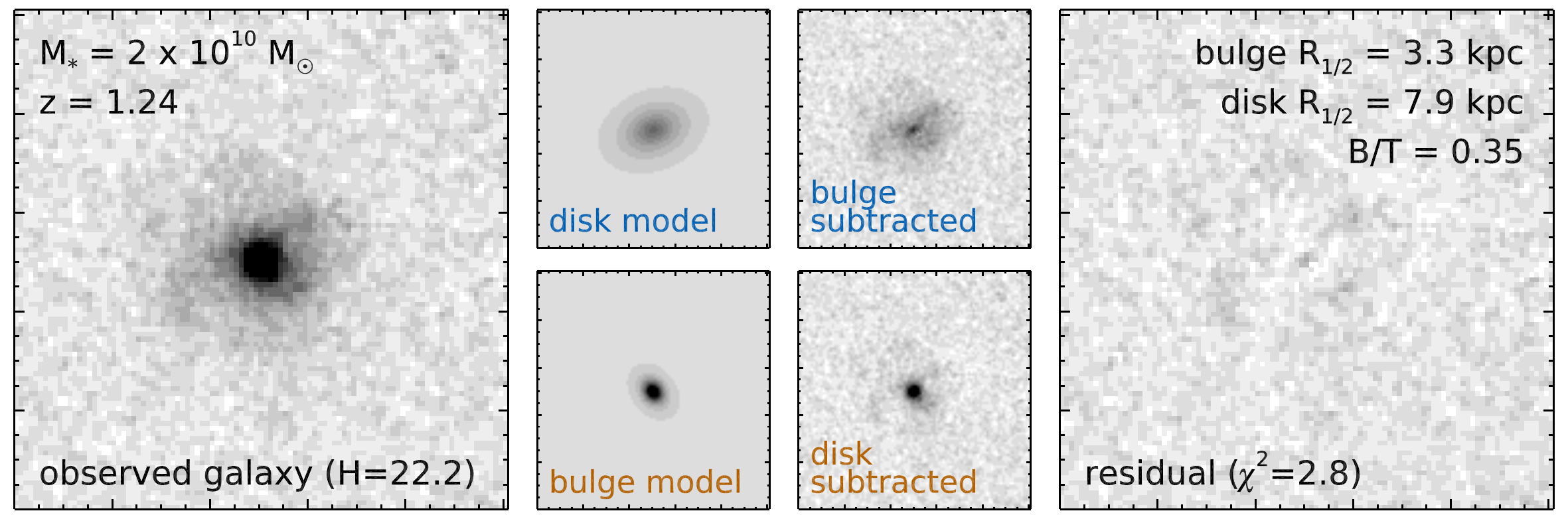}
    \caption{Example bulge-to-disk decomposition of an $H=22.2$ galaxy from the GOODS--{\it South} field, which is among the faintest galaxy in our sample. The first column shows the observed HST WFC3 image of the galaxy, and we also provide in the top-left corner its main physical properties. The second column shows the best-fit disk (top) and bulge (bottom) components as extracted by \gimtod. The third column shows the residual of the image after subtraction of the bulge (top) and the disk (bottom), to visualize the profile of the other component. Finally, the fourth column shows the residual image after both components are subtracted. The best-fit parameters are given in the top-right corner.}
    \label{FIG:g2d_fit_model}
\end{figure*}

In this section we describe the aproach we use to determine the disk stellar masses of our $z=1$ galaxies. In \rsec{SEC:bt} we detail the morphological decomposition procedure, which tell us how much of the $H$-band flux was emitted by the bulge and the disk of each galaxy. Then, in \rsec{SEC:mdisk} we show how we use this light-weighted decomposition to infer the mass-weighted $\bt$, and the disk stellar mass. We also briefly discuss the quality of our decompositions and how they compare to the literature.

\subsection{The bulge to disk decomposition \label{SEC:bt}}

To perform the bulge-to-disk decomposition, we follow \cite{pannella2009} and use the software \gimtod \citep{simard2002} on the HST $H$-band images ($0.06\arcsec/{\rm pixel}$ resolution). To carry out a proper parametric modeling of the galaxy two-dimensional light distribution, it is of fundamental importance to obtain a careful estimate of the local background level. An extended disk or the low surface brightness wings of a high \sersic index galaxy can easily fool the fitting code and hence retrieve the wrong galaxy model \citep[e.g.,][]{haeussler2007,pannella2009-a,barden2012}. In order to avoid this issue, we run \sextractor \citep{bertin1996} on the public CANDELS $H$-band images in ``cold'' mode. This allows to us to better minimize the artificial source splitting and maximize the number of pixels assigned to each object. Our newly extracted $H$-band catalog is then cross-matched to the original CANDELS photometric catalog so that every entry is assigned a redshift and a stellar mass. Less than $10\%$ of the original sample is actually not retrieved by our cold source extraction. For the most part, these are blended objects for which a bulge-to-disk decomposition would be both impractical and uncertain, and we do not consider these in the following. For every galaxy, we then we extract a cutout in both the original image and our \sextractor segmentation map, the size of which depends on the actual galaxy angular dimensions. This ensures that \gimtod is able to properly fit for the image background and recover accurate galaxy parametric modeling.

Using these image and segmentation cutouts, we fit a combination of two \sersic profiles: an exponential disk ($n=1$) and a de Vaucouleur profile ($n=4$), both convolved with the ``hybrid'' WFC3 PSFs from \cite{vanderwel2012}. An example of such decomposition in given in \rfig{FIG:g2d_fit_model}.

Although the fit settles to physically reasonable solutions in more than $95\%$ of the cases, occasionally the effective radius of either component converges to zero, meaning that the component is essentially unresolved. In this case, there is no way to disentangle an exponential disk from a de Vaucouleur profile, and this unresolved component could be either an AGN, a nuclear starburst, or just the badly-fit core-component of a bulge. Fortunately such cases are rare, so we decided to consider them as bad fits and exclude them from the following analysis.

When defining our sample, we took care to exclude close galaxy pairs that would cause blending issues (see previous section). However, while analyzing the results of the decomposition, we also found that there are a few galaxies which are not even properly deblended in the CANDELS catalogs to begin with, e.g., because the two galaxies are too close and \sextractor considered the pair as a single object. These galaxies cannot be fitted with our procedure, and typically show large $\chisqr$. To filter out these catastrophic failures, we therefore impose a maximum value of $\chisqr < 2$. This also removes remaining catastrophic fit failures, and galaxies with too irregular morphologies. This cut excludes $10\%$ of the sample\footnote{We do not further select galaxies based on their measured morphological parameters. \cite{abramson2014} only used face-on galaxies in their $z=0$ analysis (axis ratio larger than $0.8$), arguing that the decomposition is less reliable for edge-on objects. We could not find any such trend in our simulations (see \rapp{APP:simu}), and we also checked that no systematic trend emerges in the real data if we only use face-on galaxies.}.

For each galaxy that is properly fit ($2\,081$ among the $H$-sample, $872$ among the IR-sample; see dotted line on \rfig{FIG:mstar_hist}), we now have an estimation of how the $H$-band flux is distributed between the disk and the bulge. From this decomposition, we can compute a light-weighted $\bt$, and we discuss in \rsec{SEC:mdisk} how to convert this value into a mass-weighted ratio to finally obtain the stellar mass of the disk.

\subsection{Estimating the disk mass \label{SEC:mdisk}}

\begin{figure*}
    \centering
    \includegraphics[width=14cm]{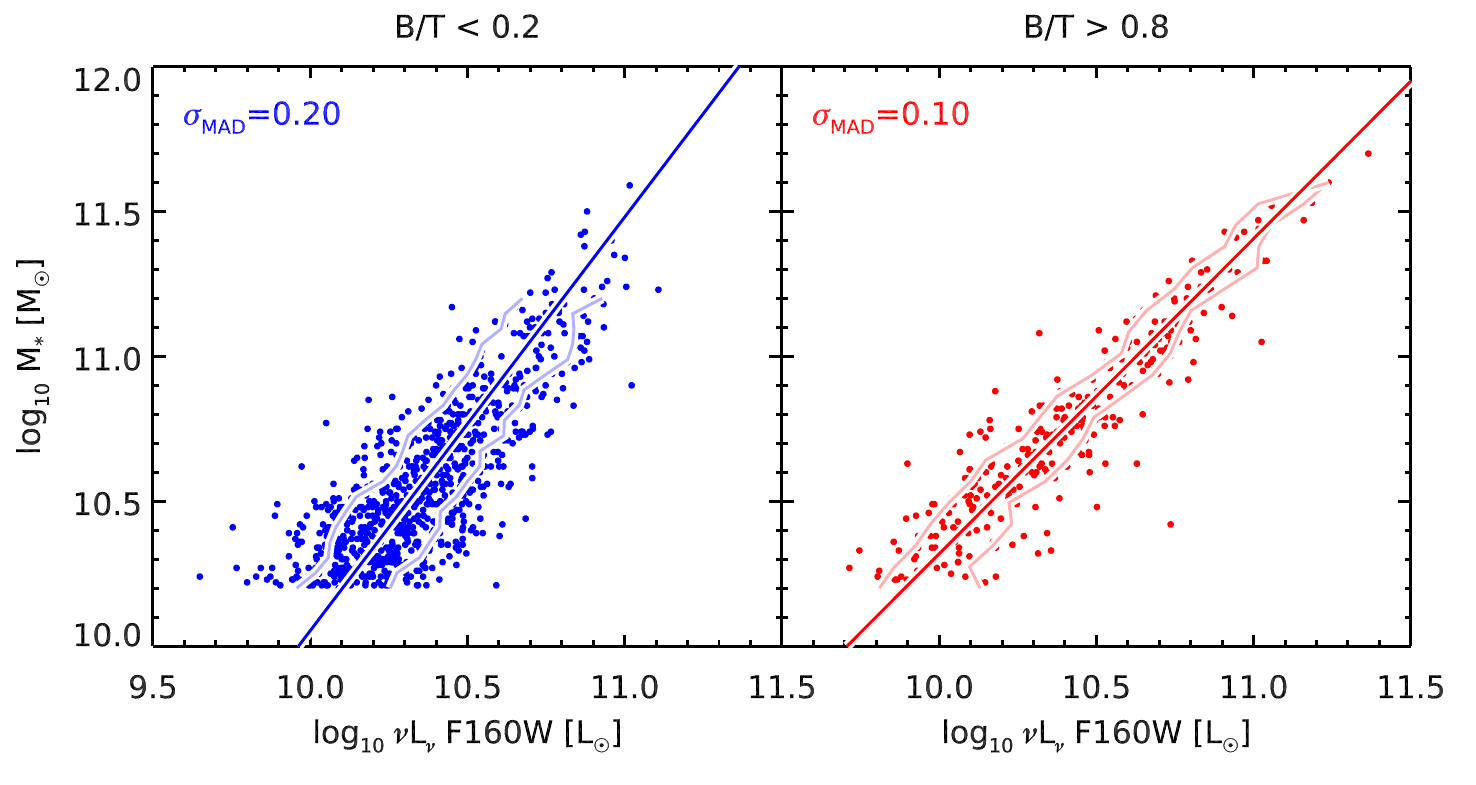}
    \caption{Relation between the total stellar mass ($\mstar$) estimated by fitting  the integrated multi-wavelength photometry of the whole galaxy and the measured luminosity from the \hst $H$-band flux (without $k$-correction) for a sample of disk-dominated galaxies ($\bt < 0.2$, left) and bulge-dominated galaxies ($\bt > 0.8$, right). Individual galaxies are shown with filled colored circles. The best-fit relation is shown with a straight line, and the dispersion around this relation is shown with light solid lines on each side. The global dispersion is given in the top-left corner of each plot, and is computed from the median absolute deviation (MAD) using $1.48\times{\rm MAD}(\Delta \mstar)$.}
    \label{FIG:ml_ratio}
\end{figure*}

\begin{figure*}
    \centering
    \includegraphics*[width=18cm]{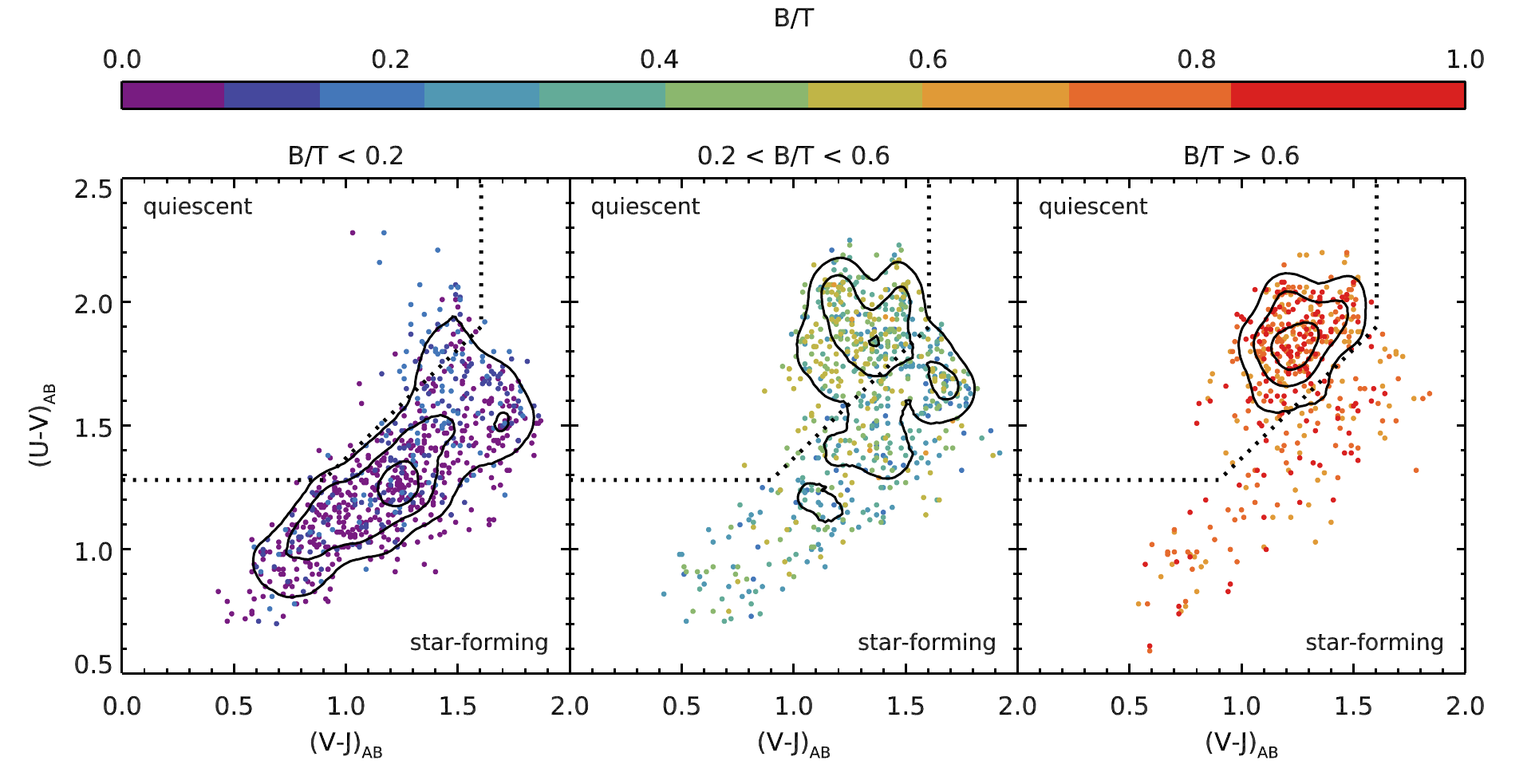}
    \caption{Location of galaxies from the $H$-sample with varying mass-weighted $\bt$ on the \uvj diagram (left: $\bt < 0.2$, middle: $0.2 < \bt < 0.6$, right: $\bt > 0.6$), using the total magnitudes of each galaxy. The dotted line shows the dividing line between the star-forming and quiescent populations defined in \req{EQ:uvj}. It is clear that both bulge- and disk-dominated galaxies occupy very different regions of the diagram, illustrating the good agreement between the colors and the morphology. Intermediate galaxies with roughly equal mass in the disk and bulge (middle panel, $\mean{\bt} = 0.4$) are spread over the two regions, with a tendency for being preferentially in the quiescent region.}
    \label{FIG:uvj_bt}
\end{figure*}

Once the flux of both the bulge and disk are measured, the last step is to measure the stellar mass of the disk. Both components have different mass-to-light ratios, since bulges are mostly made of old stars and will typically have higher mass-to-light ratios compared to the star-forming disks. In practice, since we are doing the decomposition in the $H$ band (rest-frame $i$ band at $z=1$), the variation in mass-to-light ratio is supposed to be minimal \citep[e.g.,][]{dejong1996}.

Yet, to prevent any bias in our results, we will nevertheless correct for this effect. Here we choose to follow an empirical approach where we estimate the average mass-to-light ratio for the bulge components, infer the bulge masses, and subtract them from the total stellar masses. The main advantage of this approach is that, although we perform the bulge-to-disk decomposition in a single band, we take advantage of the accurate mass-to-light ratio that was derived by fitting the total photometry of the galaxy, using a large number of photometric bands (\citetalias{schreiber2015}).

To determine the average mass-to-light ratio of bulges, we build a sample of ``pure bulge'' galaxies ($\bt > 0.8$) and compare their $1.6\,\um$ (observer frame) luminosity against the total stellar mass. Since these galaxies are clearly bulge-dominated, we can neglect the disk mass and assume that the observed mass-to-light ratio is representative of that of a bulge. The corresponding relation is shown in \rfig{FIG:ml_ratio} (right). We derive the average trend by performing a linear fit to the running median in logarithmic space and obtain
\begin{equation}
\frac{\mbulge}{\msun} = \left(\frac{\nu L_{\nu\rm, bulge}}{3.25\,\lsun}\right)^{1.09}\,,
\end{equation}
with a constant residual scatter of about $0.1\,\dex$. We then use this relation for all the other galaxies that are not bulge-dominated to estimate $\mbulge$, and subtract this value from $\mstar$ to obtain $\mdisk$.

However, we rely here on the low scatter of the mass-to-light ratio in bulges. It is true that this ratio is less variable in bulges than in star-forming disks (see, e.g., \rfig{FIG:ml_ratio}, left), because the latter can display a wider variety of star formation histories. Still, bulges are expected to show some variation of their dust content and metallicity, and this will not be taken into account here. In particular, one possibility we cannot account for is that bulges in composite or disk-dominated galaxies may have different colors than pure bulges. Lastly, another downside of this empirical approach is that, since we do not measure the colors of each individual bulge, we cannot flag out the ``blue bulges'', which are \emph{not} bulges but likely compact nuclear starbursts. These are supposed to be rare though, and if anything, this population would end up substantially above the Main Sequence in the $\sfr$--$\mdisk$ relation and bias the slope toward higher values.

In \rfig{FIG:uvj_bt}, we show on the \uvj diagram the location of galaxies that are either disk-dominated ($\bt < 0.2$), intermediate ($0.2 < \bt < 0.6$), and bulge-dominated ($\bt > 0.6$) according to our mass-weighted bulge-to-total ratios. Reassuringly, the disk-dominated galaxies populate preferentially the \uvj star-forming branch, while the bulge-dominated galaxies pile up in the quiescent cloud, although there is some overlap between the two populations close to the dividing line. Intermediate objects are preferentially in the quiescent region, but are also widely spread in the tip of the star-forming branch. This illustrates the good agreement between the morphological classification and the properties of the stellar populations, which is the high-redshift equivalent of the Hubble sequence (see also \citealt{wuyts2011}).

Lastly, it should be noted that the relations we find between total stellar mass and $\bt$ for \uvj star-forming and quiescent galaxies are consistent with those derived in \cite{lang2014}.

\section{Measuring gas masses \label{SEC:sfe}}

In this section, we describe the measurement of dust masses from the FIR to submm photometry, detailed in \rsec{SEC:mdust}, and then detail the derivation of the associated gas masses in \rsecs{SEC:metal} and \ref{SEC:mgas}.

The conversion from $\mdust$ to $\mgas$ is made using the dust-to-gas ratio, $\gdr$, which we estimate in this section. This ratio is not universal, and it is known to anti-correlate with the metallicity \citep[e.g.,][]{draine2007-a,leroy2011,sandstrom2013,remy-ruyer2014}. This anti-correlation can be simply understood if a universal fraction $f_{\rm d}$ of all the metals in the ISM are locked into dust grains, while the remaining fraction remains mixed with the gas \citep[e.g.,][]{franco1986,zafar2013}. With this assumption and a measurement of the dust mass, one just needs to know the gas-phase metallicity ($Z$) to infer the gas mass:
\begin{equation}
\mgas = \gdr\,\mdust = \frac{1}{Z}\,\frac{1-f_{\rm d}}{f_{\rm d}}\,\mdust\,. \label{EQ:mgasth}
\end{equation}
The value of $f_{\rm d}$ can be inferred empirically from observations where both the dust and the gas masses are known. In these cases, the gas mass is usually inferred by adding together $21\,{\rm cm}$ measurements of the neutral atomic hydrogen, and estimates of the molecular hydrogen mass, which are typically obtained from the carbon monoxide (CO) emission lines (since, indeed, molecular hydrogen is extremely hard to observe directly). This latter step implies yet another uncertainty on the conversion factor from CO intensity to molecular gas mass ($\alpha_{\rm CO}$). To alleviate this problem, \cite{leroy2011} performed a resolved analysis of local galaxies, inferring jointly the gas-to-dust ratio and $\alpha_{\rm CO}$ from combined dust, $\hone$ and CO observations \citep[see also][]{sandstrom2013}. Assuming that the gas-to-dust ratio remains constant throughout each galaxy, they observed the relation between $\gdr$ and metallicity, and found a dependence that is consistent with \req{EQ:mgasth}. In the present paper, we therefore use their observations to estimate $\gdr$ for all the galaxies in our sample, and therefore $\mgas$. This approach has been used extensively in the recent literature to estimate the gas masses of distant galaxies \citep[e.g.,][]{magdis2011,magnelli2012,magdis2012,santini2014,scoville2014,bethermin2015-a}.

Since most galaxies in the HRS survey have $\hone$ and CO data (at least at the high-mass end), we cross-check in \rsec{SEC:mdustgas} our dust-based gas masses by comparing them against the values obtained more straightforwardly from the $\hone$+CO measurements.

\subsection{Dust masses \label{SEC:mdust}}

\begin{figure}
    \centering
    \includegraphics[width=9cm]{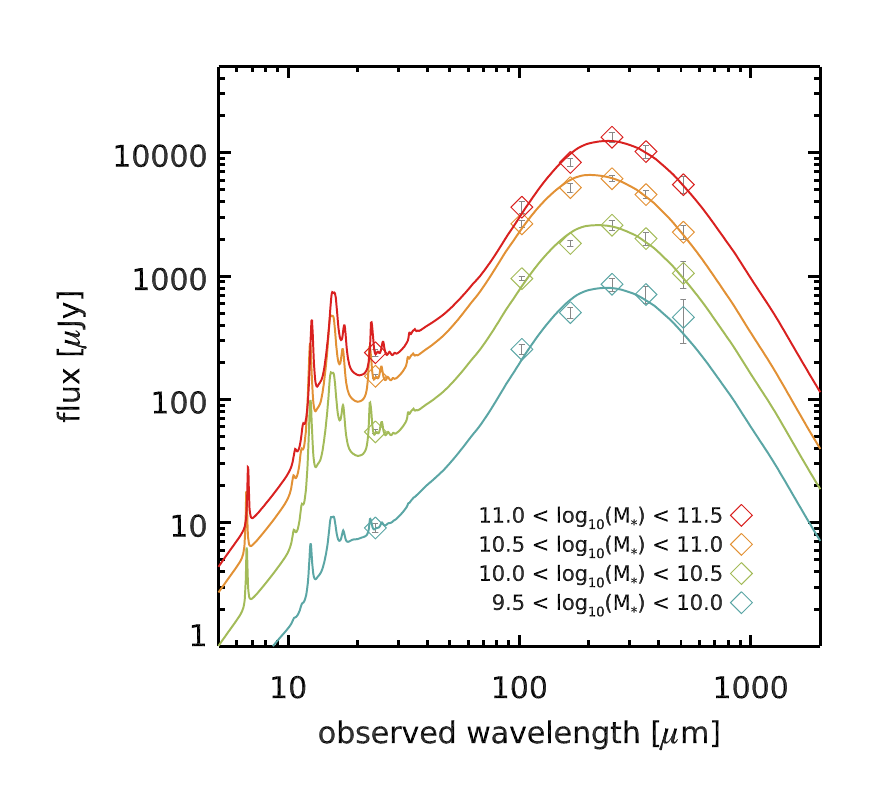}
    \caption{Mean stacked FIR SEDs of star-forming galaxies in our $z=1$ sample, split in four mass bins. The broadband photometry (open diamonds) is taken from \citetalias{schreiber2015}. The fit to the stacked measurements is performed using the dust models of \cite{galliano2011}. It is apparent from this figure that massive galaxies (in red) have a colder dust temperature. This can be seen clearly from the peak wavelength of the best-fit model, or indirectly from the flux ratio $S_{500}/S_{100}$.}
    \label{FIG:seds}
\end{figure}

\begin{table*}
\begin{center}
\begin{tabular}{ccccccccccc}
\hline \\[-0.35cm]
$\mstar$       & $\mdust$             & $\lir$               & $\tdust$             & $\fpah$             & $\sfr$               & $\logoh$ & $\mgas/\mdust$    & $\mgas$             & $\sfe$                 & $f_{\rm gas}$     \\
$10^{10}\msun$ & $10^7\msun$          & $10^{10}\lsun$       & $\kelvin$            & \%                  & $\msun/\yr$          & (PP04 [\ion{N}{ii}])   &     & $10^{10}\msun$      & $1/\Gyr$               & \%                   \\        \hline\hline \\[-0.35cm]
$0.56$         & $2.1^{+0.9}_{-0.5}$  & $2.4^{+0.2}_{-0.2}$  & $24.5^{+1.3}_{-1.4}$ & $0.8^{+0.9}_{-0.5}$ & $5.5^{+0.3}_{-0.4}$  & $8.34$   & $381^{+21}_{-25}$ & $0.8^{+0.3}_{-0.2}$ & $0.69^{+0.22}_{-0.20}$ & $58.3^{+7.7}_{-7.1}$ \\[0.1cm]
$1.8$          & $5.2^{+0.8}_{-0.5}$  & $8.7^{+0.3}_{-0.3}$  & $26.1^{+0.3}_{-0.7}$ & $4.5^{+0.2}_{-0.2}$ & $16.7^{+0.4}_{-0.5}$ & $8.48$   & $278^{+17}_{-23}$ & $1.4^{+0.3}_{-0.2}$ & $1.16^{+0.14}_{-0.16}$ & $45.0^{+4.0}_{-3.2}$ \\[0.1cm]
$5.5$          & $10.2^{+0.7}_{-0.9}$ & $23.0^{+0.9}_{-0.8}$ & $27.7^{+0.6}_{-0.5}$ & $4.9^{+0.3}_{-0.3}$ & $40.9^{+1.5}_{-1.4}$ & $8.63$   & $193^{+11}_{-13}$ & $2.0^{+0.2}_{-0.2}$ & $2.07^{+0.27}_{-0.23}$ & $26.4^{+1.9}_{-2.3}$ \\[0.1cm]
$16$           & $34.7^{+4.1}_{-3.2}$ & $41.7^{+2.3}_{-2.1}$ & $24.5^{+0.4}_{-0.5}$ & $4.4^{+0.3}_{-0.3}$ & $73.3^{+3.8}_{-3.7}$ & $8.76$   & $145^{+9}_{-6}$   & $5.0^{+0.7}_{-0.4}$ & $1.45^{+0.15}_{-0.19}$ & $24.7^{+2.4}_{-2.1}$ \\[0.1cm]
\hline
\end{tabular}
\end{center}
\caption{\label{TAB:stackprop} Average physical properties of the galaxies in the stacked $z=1$ sample. The quoted errors indicate the uncertainty on the average, not the intrinsic spread of the population. These uncertainties are derived through bootstrapping half of the full sample, recomputing all quantities for each bootstrap realization separately, then measuring the standard deviation among all realizations. The gas-to-dust ratio is randomized within the allowed statistical uncertainty (\req{EQ:gdr}). The resulting values are then divided by $\sqrt2$ to take into account that only half of the initial sample was used in each bootstrap realization.}
\end{table*}

Accurate dust masses can only be derived from FIR measurements down the Rayleigh-Jeans tail of the dust continuum, meaning at $z=1$ that we need to measure the observer-frame emission of galaxies at $\lambda \ge 400\,\um$. While \herschel does provide deep imaging at $500\,\um$, the poor angular resolution prevents measuring the $500\,\um$ flux of most galaxies, since finding the right counterpart to the fluxes measured on these maps is challenging (see, e.g., \citealt{shu2015}).

This issue can be avoided by stacking the images, since the contribution from neighboring sources averages out to form a constant background. However, if galaxies tend to be clustered on the sky, the contribution of neighboring sources will \emph{not} average out to a strictly uniform value, and will instead tend to produce more flux toward to the position of the stacked galaxies \citep[see, e.g.,][]{bethermin2010}. This is particularly important for the present study, since the amplitude of this effect will depend on the size of the beam, and will therefore affect preferentially the longest wavelengths which are the ones that best correlate with the dust mass. In \citetalias{schreiber2015}, we implemented an empirical correction to remove this flux boosting, which was derived from a set of realistic simulated images. The stacked $500\,\um$ fluxes in the simulation were found to be boosted by $20\%$ on average, and we therefore corrected the observed fluxes by that same amount. After this factor is taken into account, no remaining bias was found in the stacked fluxes\footnote{To better constrain the Rayleigh-Jeans tail of the dust emission, we also considered stacking longer wavelength sub-millimeter data from AzTEC or LABOCA, however these are only available for a few fields (AzTEC in GOODS--{\it North} and LABOCA in GOODS--{\it South}, while both are also covering COSMOS at shallower depth) hence reducing significantly the number of stacked sources. Combined with the fact that, at $z=1$, the expected flux in these bands is fairly low, we could not detect any significant signal. These upper limits are consistent with the rest of \herschel photometry at the $1$ to $2\sigma$ level.}.

For our $z=1$ sample, we therefore use the stacked SEDs of \citetalias{schreiber2015}, which are reproduced here in \rfig{FIG:seds}. These SEDs were built by stacking all the \uvj star-forming galaxies in the four CANDELS fields at $0.7 < z < 1.3$ and in four bins of stellar mass: $\log_{10}(\mstar/\msun) = 9.5$ to $10$, $10$ to $10.5$, $10.5$ to $11$ and $11$ to $11.5$. As described above, a correction for clustering is also applied.

We then analyze the stacked FIR photometry with a library of template SEDs built from the amorphous carbon dust model of \cite{galliano2011}. This new library will be presented in a forthcoming paper (Schreiber et al.~in prep.), and is introduced to extend the \cite{chary2001} SED library (hereafter CE01) with the aim to provide a wider and finer grained range of dust temperatures (or, equivalently, $\lir/\mdust$) and finer control on the PAH mass-fraction (or, equivalently, $\ireight\equiv\lir/\leight$).

We fit the stacked \herschel photometry with each template of the library, corresponding each to a different value of $\tdust$ (or $\mean{U}$), and pick the one that best fits the observed data. Essentially, there is a direct mapping between the dust temperature and the position of the peak of the FIR emission (i.e., Wein's law): SEDs peaking at longer wavelengths (which is the case of our highest mass bin) have lower dust temperatures. Then, since each SED in the library is calibrated per unit $\mdust$, the dust mass is trivially obtained from the normalization of the best-fit template. Here, we allow the dust temperature to vary between $15$ and $50\,{\rm K}$, while the PAH mass-fraction is left free to vary between $0$ and $1$.

The best-fit values we obtain are referenced in \rtab{TAB:stackprop}, and the best-fit models are shown in \rfig{FIG:seds}. While our models accurately describe the observed data, we find a systematic offset of the order of $20\%$ in the PACS bands, where the $100\,\um$ and $160\,\um$ fluxes are respectively above and below our model. No such trend is found for the three SPIRE bands. These offsets could be caused partly by calibration uncertainty (of the order of $15\%$ for \herschel; \citealt{poglitsch2010,swinyard2010}), but also by the limited number of free parameters in our dust models\footnote{E.g., we could improve the fit by adopting overall lower dust temperatures and adding a second component of warm dust as in \cite{dacunha2008}.}. However, these offsets are small and affect all mass bins in a similar way; they will therefore not impact our results.

For galaxies in the HRS, angular resolution is not an issue and the \herschel photometry of each galaxy can be obtained and fitted individually without stacking. The dust masses are estimated exactly as for our stacked $z=1$ SEDs, fitting the mid- to far-IR SED of the individual HRS galaxies with our template SED library. More detail on the IR photometry and dust properties of these objects is given in \cite{ciesla2014}\footnote{We would reach the same conclusions had we used the dust masses published by Ciesla et al., after correcting them downward by a factor of $2$ since these were derived using the \cite{draine2007} graphite dust model.}.

As a cross check, we also fit the FIR photometry with the CIGALE SED fitting code, using the \cite{draine2007} dust SED library. While we recover identical $\lir$, the $\mdust$ values obtained with the \cite{draine2007} models are systematically higher by a factor of two compared to our own estimates. Systematic differences in the dust masses are typically found by comparing the results of two different approaches, e.g., comparing the results from the \cite{draine2007} library against simple modified black bodies (as is shown in \citealt{magdis2012} and \citealt{magnelli2012}), or different chemical compositions of dust grains within the same model \citep[e.g., graphite and silicate {\it versus} amorphous carbon grains, as in][]{galliano2011,remy-ruyer2015}. The factor of two we observe here is consistent with the value reported by \cite{galliano2011}, who argue that dust masses derived by models using graphite \cite[like, e.g., the models of][]{draine2007} instead of amorphous carbon grains are overestimated by a factor of $2.6$. They also claim that this overestimation creates a tension with the measured metallicity of the Large Magellanic Cloud by violating the element abundances, and therefore advocate instead the use of amorphous carbon grains in dust models. Similar conclusions have been drawn in the Milky Way and other nearby galaxies \citep{compiegne2011,jones2013,planckcollaboration2014,fanciullo2015}.

This emphasizes that, without precise knowledge of the detailed chemical composition of the dust, the \emph{absolute} value of the dust masses should be taken with a grain of salt. Since we are only interested in the \emph{relative} evolution of the gas mass with stellar mass in this work, this issue is of no consequence \emph{provided that galaxies of different stellar masses host dust grains of similar chemical composition}. The latter is a key assumption of our approach. In the Local Universe, the properties and composition of the dust are known to vary, in particular as a function of metallicity \citep{madden2006,wu2006,ohalloran2006,smith2007,draine2007-a,galliano2008-a,ciesla2014,remy-ruyer2015}. However, since our samples are composed mostly of galaxies with close-to-solar metallicity (at least $0.4\,Z_{\odot}$ in both our $z=0$ and $z=1$ samples, see next section), we do not expect our galaxies to exhibit strong variations of their dust composition. In \rsec{SEC:mdustgas}, we nevertheless check that this assumption holds by comparing our dust-based gas masses against more direct measurements from $\hone$ and CO measurements in the HRS.

\subsection{Metallicities \label{SEC:metal}}

Once the dust masses are measured (see previous section), the next step toward the determination of the gas masses is to estimate the metallicity. Since only half of the galaxies in the HRS have individual metallicity measurements \citep{hughes2013}, and almost none of the galaxies in our $z=1$ sample, we need to use empirical recipes to estimate the metallicities. Following recent literature \citep[e.g.,][]{magdis2012,santini2014,bethermin2015-a}, we estimate the metallicity from the Fundamental Metallicity Relation \citep[FMR,][Eq.~5]{mannucci2010}
\begin{align}
&(\logoh)_{\rm KD02} \nonumber \\
&\quad\quad = \left\{\begin{array}{lcl}
    8.9 + 0.47\,(\mu_{0.32} - 10) & \text{for} & \mu_{0.32} < 10.36 \\
    9.07 & \text{for} & \mu_{0.32} \geq 10.36 \\
\end{array}\right., \label{EQ:fmr}
\end{align}
with $\mu_{0.32} \equiv \log_{10}(\mstar\,[\msun]) - 0.32 \times \log_{10}(\sfr\,[\msun/\yr])$, and where both $\mstar$ and $\sfr$ are converted to the \cite{chabrier2003} IMF (i.e., divided by $1.67$ from the Salpeter values, as in \citealt{madau2014}). For our $z=1$ sample, we use the average stellar mass and $\sfr$ obtained in the stacks (see previous section), and for the $z=0$ HRS galaxies without metallicity measurement we use their respective $\mstar$ and $\sfr$. We checked that using this prescription or estimating the metallicity from the $z=1$ mass--metallicity relation \citep[e.g.,][]{zahid2011} would not change our conclusions ($+0.12\,\dex$ metallicity shift at $z=1$, after accounting for the different calibration\footnote{It is also worth noting that the FMR could have a redshift-dependence, i.e., that \req{EQ:fmr} may not hold in the distant Universe \citep[see in particular][]{troncoso2014,tan2014,bethermin2015-a}. However, this is not an issue for the present study since, first, this difference is supposedly a constant shift of the metallicity at all stellar masses, and second, it only takes place at higher redshifts than that probed by our study.}).

On the other hand, \cite{kewley2008} showed that there exists substantial systematic differences of metallicity measurements, depending both on the available observables used to derive the oxygen abundance, and the calibration that is used. For example, the FMR was derived using the \cite{kewley2002} (KD02) calibration, while the metallicities of \cite{magdis2012} are obtained with the prescription of \cite{pettini2004} (PP04). According to \cite{kewley2008}, the difference between these two metallicity estimates is roughly constant and equal to about $0.25\,\dex$ (at least in the metallicity range considered in this paper), with a scatter of only $0.05\,\dex$: it is only a global shift of the absolute metallicity, and will not affect the relative trends. To derive accurate dust-to-gas ratios, it is nevertheless important to make sure that the same metallicity calibration is used consistently in all calculations.

In the following section, we derive a relation between the gas-to-dust ratio and the metallicity, assuming the metallicity is given in the ``PP04 [\ion{N}{ii}]'' scale. To use this relation, we therefore need to convert the FMR metallicities derived above to this new scale, which we do following the calibration proposed by \cite{kewley2008}:
\begin{align}
(\logoh)_{\rm PP04} &= 569.4927 - 192.5182\,x \nonumber \\
&+ 21.91836\,x^2 - 0.827884\,x^3\,,
\end{align}
with $x \equiv (\logoh)_{\rm KD02}$. As written above, in practice for the galaxies we consider in this study these ``PP04'' abundances are systematically lower by $0.3\,\dex$ compared to the original ``KD02'' values (this constant shift holds within $0.05\,\dex$ for all $\logoh_{\rm KD02} > 8.5$).

The measured metallicities of the HRS galaxies are already in this scale, and needed no conversion. For HRS galaxies with a metallicity measurement, comparing the latter to the metallicity derived from the FMR, we find a median offset of $0.08\,\dex$ and a scatter of $0.1\,\dex$, consistent with the values reported by \cite{mannucci2010}. Since these latter values are low, and to avoid mixing together metallicities that are directly observed and those that are inferred from the FMR, we decide to use the FMR-based metallicities for all galaxies in the HRS. We checked that our results are not affected by this choice. Furthermore, the low scatter we observe in this comparison confirms the accuracy of the FMR in determining metallicities empirically. While the scatter of the FMR could increase toward higher redshifts, it should be noted that our $z=1$ stacked measurements are not sensitive to this scatter, since we only consider the average properties of galaxy populations with similar stellar masses, for which the FMR will give an accurate estimate of the average metallicity by construction.

\subsection{Gas-to-dust ratios and gas masses \label{SEC:mgas}}

The last step to estimate gas masses is to derive the gas-to-dust ratios. To do so, we employ \req{EQ:mgasth} that we calibrate using the $\gdr$ measured in a sample of local galaxies by \cite{leroy2011} (using the revised ``PP04''metallicities from \citealt{magdis2012}), that we multiply by a factor of $2$ to account for systematic differences in the dust mass measurements between the dust model that we used and that of \cite{draine2007} (see \rsec{SEC:mdust}). Assuming the linear metallicity dependence of \req{EQ:mgasth}, we find that the $\gdr$ measured by \cite{leroy2011} are well described by
\begin{align}
\log_{10}\left(\gdr\right) &= (10.92 \pm 0.04) - (\logoh)_{\rm PP04}\,, \label{EQ:gdr}
\end{align}
With a solar oxygen abundance of $(\logoh)_{\sun} = 8.73\pm0.05$ \citep{asplund2009}, this leads to the equivalent expression
\begin{align}
\gdr = (155 \pm 23)\times\frac{Z_{\sun}}{Z}\,, \label{EQ:gdr2}
\end{align}
which is consistent with the gas-to-dust ratio of the Milky Way $(\mgas/\mdust)_{\rm MW} = 158$ \citep{zubko2004}. Coming back to \req{EQ:mgasth}, using a solar metallicity of $Z_{\sun} = 0.0134$ (\citealt{asplund2009}, assuming an uncertainty of $0.001$) we note that this prescription is therefore equivalent to assuming $f_{\rm d}=(32\pm4)\%$, which is below the maximum value of $\sim46\%$ allowed by the observed metal depletion of the ISM in the Milky Way\footnote{Using the dust masses from the Draine \& Li models would increase our estimate of $f_{\rm d}$ to $51\%$.} (e.g., \citealt{draine2007-a}).

For our $z=1$ sample, \req{EQ:gdr} (or \req{EQ:gdr2}) yields gas-to-dust ratios between $145$ and $381$ (the precise values we obtain are listed in \rtab{TAB:stackprop}), while it ranges from $145$ to $488$ for the $z=0$ HRS galaxies (which cover a wider metallicity range). Using the dust masses we measured in \rsec{SEC:mdust}, we can infer the total gas mass in each stacked bin at $z=1$, and for each HRS galaxy.

\subsection{Evaluation of dust-based gas mass estimates \label{SEC:mdustgas}}

\begin{figure*}
    \centering
    \includegraphics*[width=16cm]{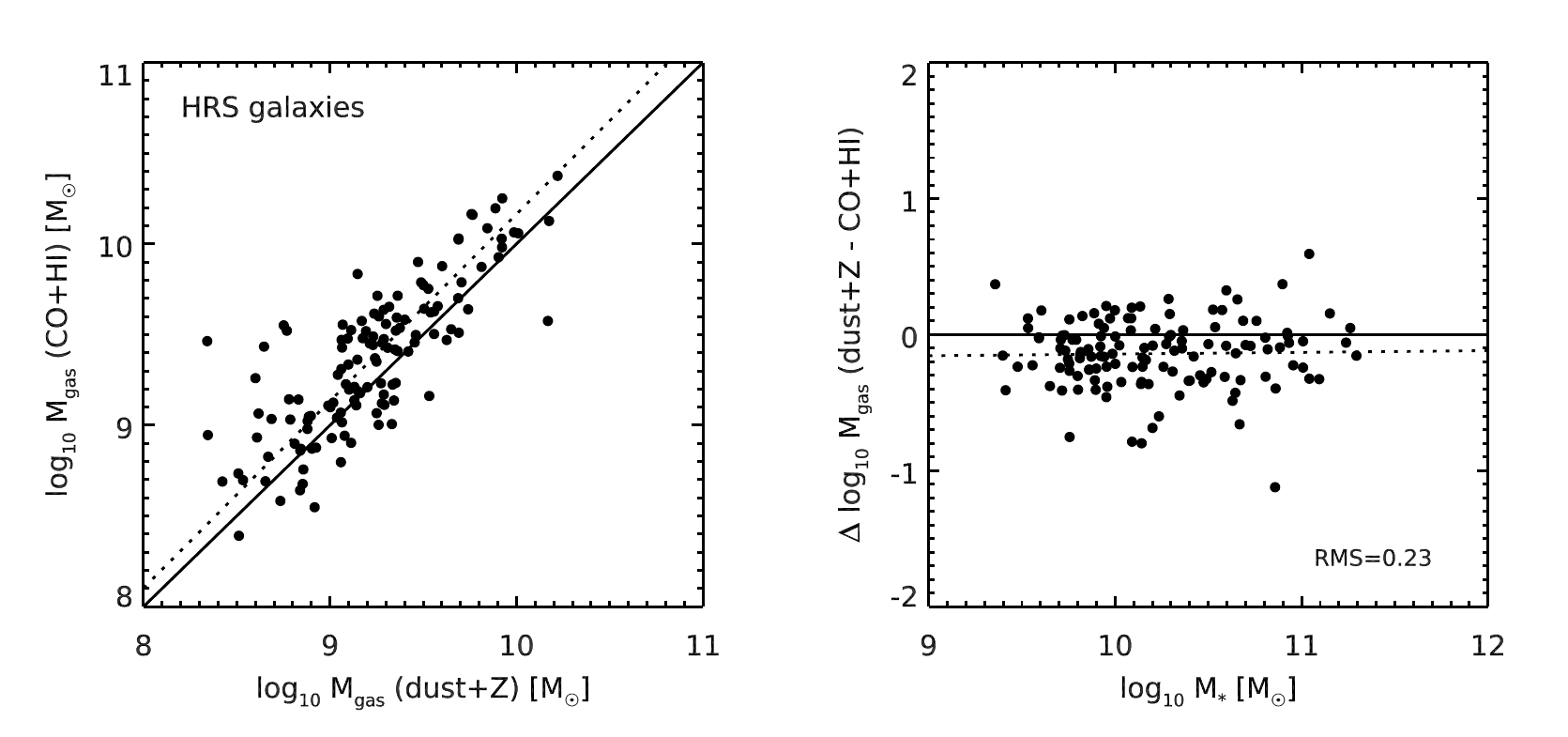}
    \caption{{\bf Left:} Comparison of two independent estimates of the total ($\hone+\htwo$) gas masses for the HRS galaxies, either using the dust mass and the metallicity as described in \rsec{SEC:mgas} (x-axis) or using a more direct measurement from $\hone$+CO spectroscopy (y-axis). The black solid line shows the one-to-one relation, while the dotted line gives the best-fit linear trend (slope: $1.03 \pm 0.03$). {\bf Right:} Difference between these two independent gas mass estimates as a function of stellar mass. The black solid line is the line of perfect agreement, while the dotted line is the best-fit linear trend (slope: $0.01 \pm 0.04$).}
    \label{FIG:mdustgas}
\end{figure*}

The procedure described above involves many steps with potential uncertainties and biases. While each of these steps has already been calibrated in the literature, it remains to check that the overall procedure (which essentially boils down to estimating gas masses from dust masses, stellar masses and star formation rates) works correctly. We can do so using the exquisite data set from the HRS. Indeed, since a substantial fraction of the galaxies in this sample are covered with $\hone$ and CO surveys \citep{boselli2014-a}, we can directly compare our dust-based gas masses against the $\hone$+CO values, assuming a constant $\alpha_{\rm CO} = 3.6\,\msun/(\kelvin\,{\rm km}/({\rm s}\,{\rm pc}^2))$ \citep{strong1988} to derive molecular gas masses.

The result is shown in \rfig{FIG:mdustgas}, either comparing the two gas mass estimates directly (left), or as a function of stellar mass (right). The $\hone$+CO gas masses are found to be systematically larger by $30\%$, and with a scatter of $0.2\,\dex$. The data do not indicate any significant differential trend with stellar mass; we find a potential bias of only $(5\pm14)\%$ between our two extreme mass bins. Since the vast majority ($90\%$) of the $\mstar > 10^{10}\,\msun$ star-forming galaxies are detected in both atomic and molecular surveys, we also perform the analysis of the next sections with these alternative gas mass estimates. We find that our conclusions remain unchanged, save for this global shift of the gas masses by a factor of $1.3$, and therefore conclude that our dust-based gas mass estimates in the HRS are robust. Since our $z=1$ sample probes a more limited metallicity range (owing to its higher stellar mass cut), we can safely assume that the same conclusion holds for this sample as well.

\section{Results \label{SEC:results}}

\begin{figure*}
    \centering
    \includegraphics*[width=18cm]{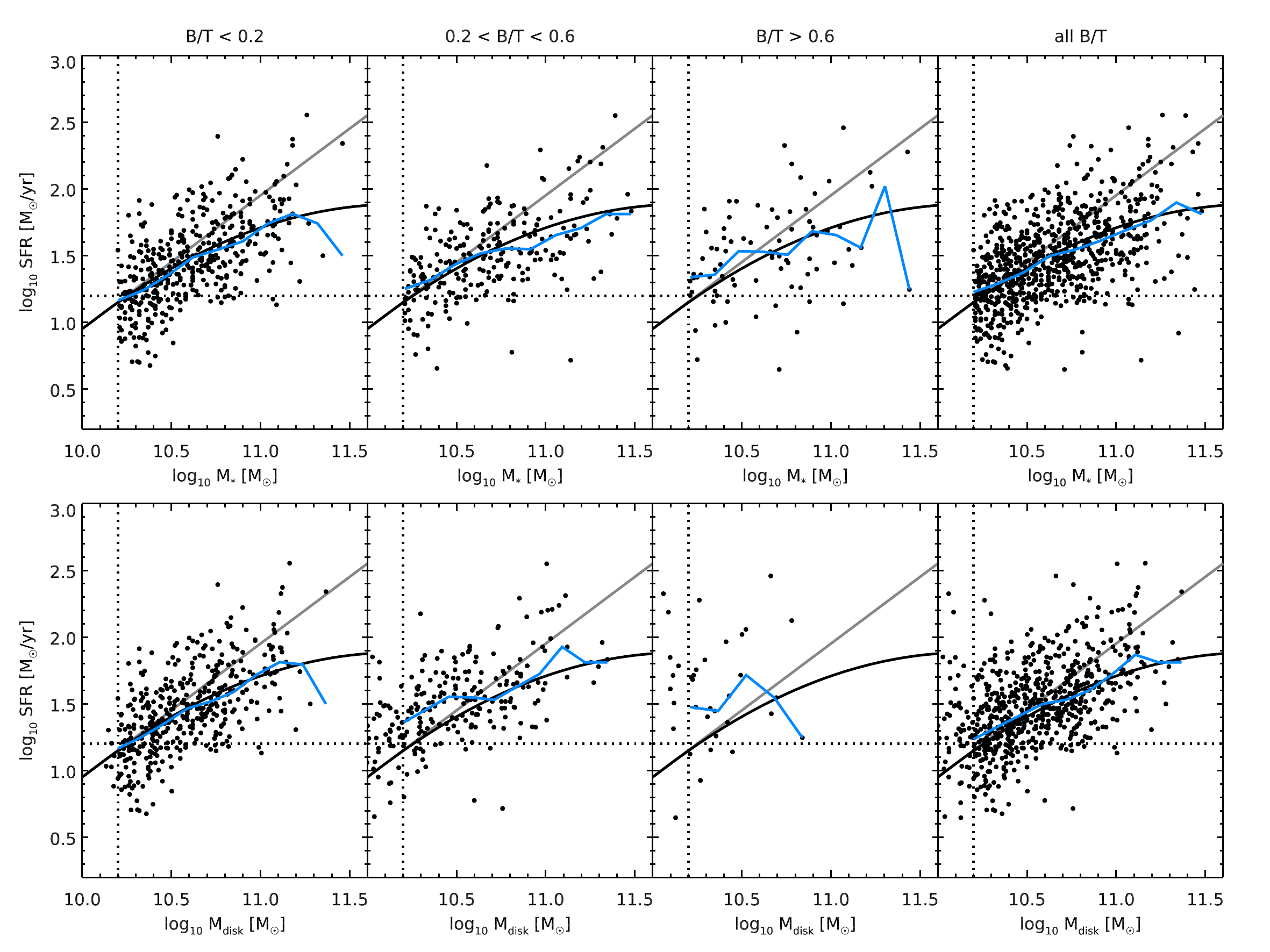}
    \caption{{\bf Upper panel:} Location of galaxies from the IR-sample with varying $\bt$ on the $\sfr$--$\mstar$ plane, using the stellar mass and star formation rate (IR+UV) of the whole galaxy. On all plots, the vertical dotted line shows our adopted stellar mass cut, the horizontal dotted line is the $90\%$ completeness in $\sfr$, and the solid black line shows the locus of the $z=1$ Main Sequence as observed through stacking in \citetalias{schreiber2015}, while the solid gray line shows the extrapolation of the low-mass trend assuming a slope of unity, as observed at lower stellar masses (see \rfig{FIG:sfrms_sample}). In each column, galaxies of different $\bt$ are plotted. In the rightmost panel, we show all galaxies regardless of their $\bt$. The solid blue lines show the running median of the sample. {\bf Lower panel:} Same as upper panel, but on the $\sfr$--$\mdisk$ plane.}
    \label{FIG:sfrms_bt}
\end{figure*}

\begin{figure*}
    \centering
    \includegraphics*[width=14cm]{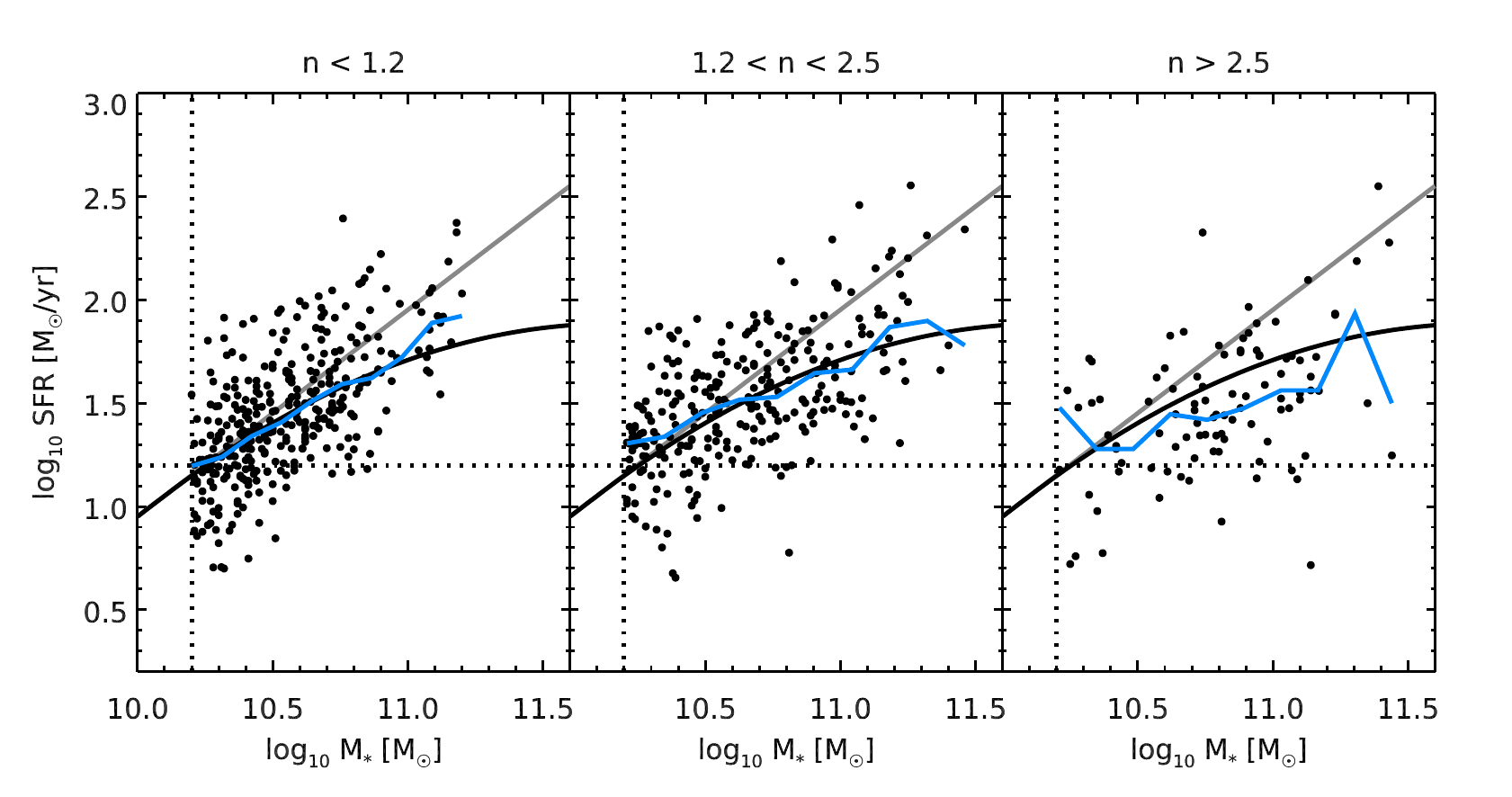}
    \caption{Same as the upper panel of \rfig{FIG:sfrms_bt}, but this time varying the \sersic index $n$.}
    \label{FIG:sfrms_index}
\end{figure*}

\subsection{The $\sfr$--$\mdisk$ relation at $z=1$ \label{SEC:sfrmdisk}}

\begin{table}
\begin{center}
\begin{tabular}{ccccc}
\hline \\[-0.35cm]
                 & all $\bt$     & $\bt < 0.2$   & $n < 1.2$ \\ \hline\hline \\[-0.35cm]
$\sfr$--$\mstar$ & $0.54\pm0.05$ & $0.67\pm0.07$ & $0.75\pm0.05$ \\
$\sfr$--$\mdisk$ & $0.60\pm0.05$ & $0.65\pm0.08$ & -- \\
\hline
\end{tabular}
\end{center}
\caption{\label{TAB:slopes} Measured slopes of the $\sfr$--$X$ relation, where $X$ is either $\mstar$ or $\mdisk$. All slopes were obtained by fitting a straight line (in logarithmic space) to the running median shown in Figs.~\ref{FIG:sfrms_bt} and \ref{FIG:sfrms_index}, considering only star-forming galaxies with $10.2 < \log_{10}(X) < 11.3$. Uncertainties are estimated by bootstrapping.}
\end{table}

\begin{figure*}
    \centering
    \includegraphics[width=9cm]{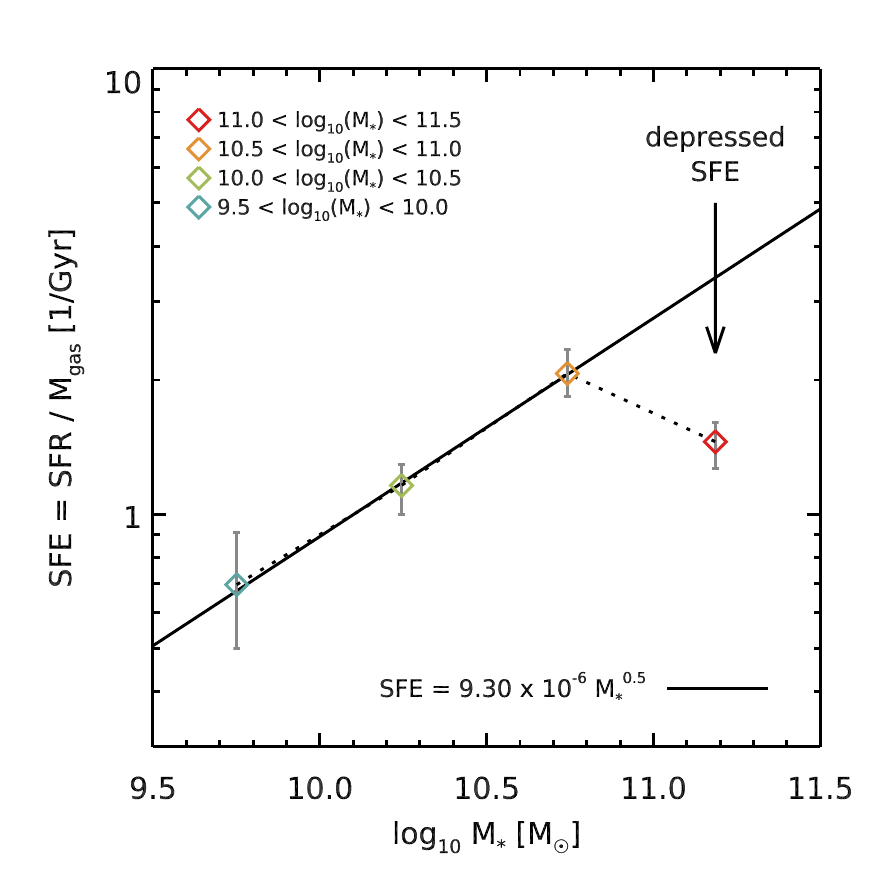}
    \includegraphics[width=9cm]{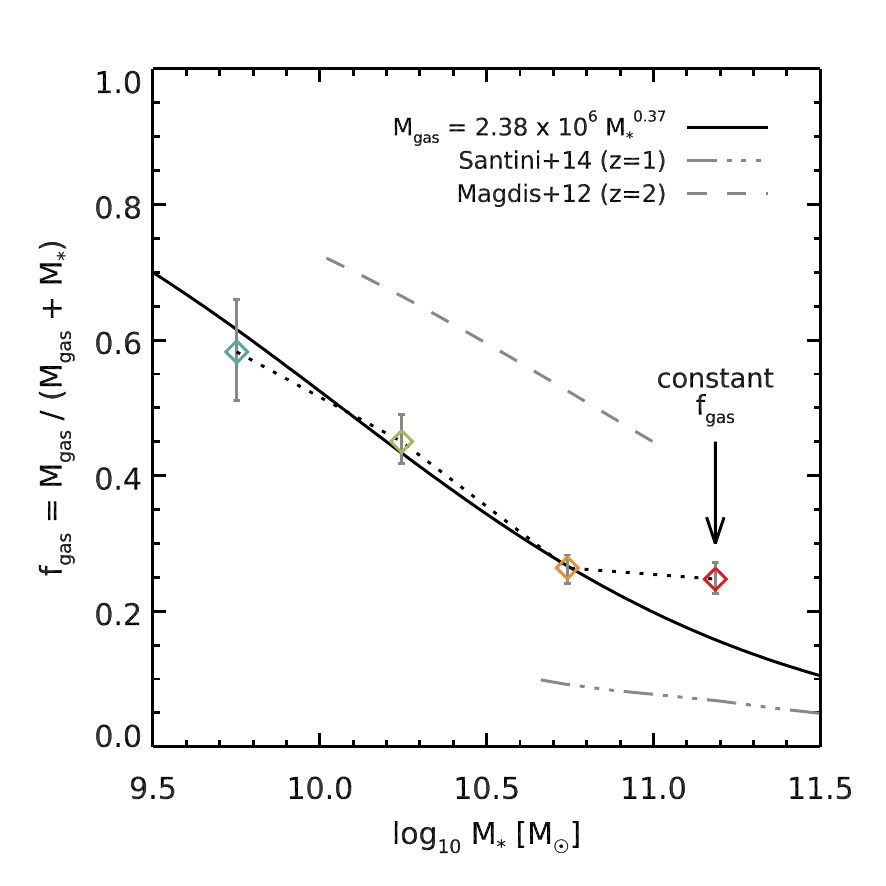}
    \caption{{\bf Left:} Relation between the $\sfe=\sfr/\mgas$ and the stellar mass ($\mstar$) for Main Sequence galaxies at $z=1$. Colored diamonds show the measured $\sfe$s and $\mstar$ of our stacked sample, the color being associated to the stellar mass as in \rfig{FIG:seds}. The best-fit power law to our measurements, excluding the most massive point, is given with a black solid line (\req{EQ:sfe}). {\bf Right:} Gas fraction ($\fgas \equiv \mgas/(\mgas + \mstar)$) as a function of the stellar mass ($\mstar$) for Main Sequence galaxies at $z=1$. The legend is the same as in the left figure, and here the solid black line gives the value of $\fgas$ computed using the best-fit $\mgas$--$\mstar$ relation, also excluding the most massive point in the fit. The resulting expression of $\fgas$ is given in \req{EQ:fgas}. We also show the measured gas fractions by \cite{magdis2012} at $z=2$ with a dashed gray line, and \cite{santini2014} at $z=1$ with a dot-dashed gray line.}
    \label{FIG:sfe}
\end{figure*}

Having measured the disk masses, we can now see if the $\sfr$--$\mdisk$ relation is universal and linear by comparing the slopes of the Main Sequence using either the total stellar mass $\mstar$ or the disk mass $\mdisk$. To be able to measure this slope on our whole sample at once, and because our redshift window is relatively large, we correct for the redshift evolution of the Main Sequence by renormalizing the $\sfr$ of each galaxy to a common redshift of $z=1$. To do so, we use the redshift evolution measured in \citetalias{schreiber2015}, taking the trend of low-mass galaxies where the bending of the Main Sequence is negligible. This correction is typically of the order of $0.05\,\dex$, and no more than $0.1\,\dex$.

In \rfig{FIG:sfrms_bt}, we show the resulting $\sfr$--$\mstar$ (top) and $\sfr$--$\mdisk$ (bottom) relations of our sample. Each panel focuses on a different range of $\bt$, starting from disks-dominated galaxies on the left, then increasing progressively the contribution of the bulge. In the rightmost panels, we show all galaxies from the IR-sample regardless of their $\bt$. We show with blue lines the running medians on the measurements in each plot, and compare them to the stacked Main Sequence of \citetalias{schreiber2015}. In the top-rightmost panel, this running median overlaps with the stacked relation, which indicates that we are not strongly affected by the $\sfr$ selection of our sample. However, we can see from the top-leftmost panel that disk-dominated galaxies do not populate a particularly different region of the $\sfr$--$\mstar$ diagram: they cluster around the stacked relation of \citetalias{schreiber2015}, and follow a sequence of slope $0.67\pm0.07$ (from $\mstar = 3\times10^{10}$ to $3\times10^{11}\,\msun$). Even after subtracting the bulge mass, which is by definition very low in these systems, the measured slope is $0.65\pm0.08$, i.e., clearly not unity. For the other galaxies, we do find a trend for some of the lowest $\ssfr$ objects to be brought back toward the Main Sequence by removing the bulge mass, but they constitute a very small fraction of the whole sample (in fact, as can be seen in \rfig{FIG:uvj_bt}, a good fraction of the bulge-dominated galaxies are classified as \uvj quiescent), and cannot counterbalance the bending observed in disk-dominated galaxies. In the end, the slope of the $\sfr$--$\mdisk$ relation as measured on the whole sample (bottom-rightmost panel) is $0.60\pm0.05$. Therefore, knowing that the Main Sequence slope at $\mstar < 10^{10}\,\msun$ is unity, we do not find that the $\sfr$--$\mdisk$ relation is linear.

In their $z=0$ study, \cite{abramson2014} only considered galaxies with $\bt < 0.6$, arguing that galaxies above this threshold cannot be fitted reliably (we show indeed in \rapp{APP:simu} that disk masses measured in bulge-dominated galaxies are the most uncertain). We therefore tried to reject galaxies with $\bt > 0.6$, and did not find any significant difference. Most of them do not show any measurable IR emission ($83\%$, compared to $46\%$ for galaxies with $\bt < 0.6$), and are likely genuine bulge-dominated and quiescent objects.

To make sure that our results are not caused by an uncertain bulge-to-disk decomposition, we show in \rfig{FIG:sfrms_index} how the $\sfr$--$\mstar$ diagram is populated by galaxies of varying effective \sersic index $n$ \cite[][and our own fits in GOODS--{\it North}, see \rsec{SEC:bt}]{vanderwel2012}. While the \sersic index alone is not well suited for measuring the disk masses of composite systems, it is a robust way of identifying disk-dominated galaxies. Indeed, the fit is intrinsically simpler and therefore more stable, and the presence of a significant bulge component will rapidly make the effective \sersic index depart from $1$, the nominal value for pure disks \cite[see, e.g., the Appendix A of][]{lang2014}. We find that disk-dominated galaxies ($n<1.2$) follow a slightly steeper slope of $0.75\pm0.05$, consistent with that found in \cite{salmi2012} and \cite{whitaker2015}, but this is still not unity. These slope measurements are summarized in \rtab{TAB:slopes}.

\subsection{Gas fraction and star formation efficiency at $z=1$ \label{SEC:sfemstar}}

We show in \rfig{FIG:sfe} (left) the behavior of the $\sfe$ as a function of the stellar mass in our stacked $z=1$ sample. These values are also reported in \rtab{TAB:stackprop}. From this figure, one can see that the $\sfe$ of galaxies at $\mstar < 10^{11}\,\msun$ rises steadily with stellar mass, following
\begin{equation}
\sfe\,[1/\Gyr] = \frac{\sfr}{\mgas} = 9.30 \times 10^{-6}\,\left(\frac{\mstar}{\msun}\right)^{0.5}\,. \label{EQ:sfe}
\end{equation}
However, our data point with the highest gas mass, i.e., corresponding to the stellar mass of $2\times10^{11}\,\msun$ where the bending of the Main Sequence is most pronounced, has an $\sfe$ that is a factor of $2$ lower than that predicted from this scaling law. Our data clearly favor two regimes of $\sfe$: low stellar mass galaxies follow a universal relation, and high stellar mass galaxies drop below this trend. Note that, owing to the uncertainty on the fiducial trend given above, we cannot rule out a weak drop of $\sfe$ in the intermediate mass bin, at $\mstar \sim 5\times10^{10}\,\msun$ (orange point).

In contrast, the gas fraction (\rfig{FIG:sfe}, right) is found to decrease continuously with stellar mass (similarly to what was found in \citealt{magdis2012} and \citealt{santini2014}). This is the expected behavior if the Main Sequence has a linear (or sublinear) slope while the $\sfr$--$\mgas$ law (the so-called integrated Schmidt--Kennicutt law) is superlinear with a power-law slope of $n > 1$ \citep[e.g.,][]{daddi2010,sargent2014,santini2014}. Indeed, if $\sfr \sim \mstar$ and $\sfr \sim \mgas^n$, then $\mgas \sim \mstar^{1/n}$ and the gas fraction has to decrease with stellar mass. By fitting the $\mgas$--$\mstar$ relation for galaxies with $\mstar < 10^{11}\,\msun$, we get
\begin{align}
\frac{\mgas}{\msun} &= 2.38\times10^{6}\,\left(\frac{\mstar}{\msun}\right)^{0.37}\,, \nonumber \\
\fgas &= \frac{\mgas}{\mgas + \mstar} = \frac{1}{1 + \left(\frac{\mstar}{1.32\times10^{10}\,\msun}\right)^{0.63}}\,. \label{EQ:fgas}
\end{align}
For galaxies with $\mstar > 3\times10^{10}\,\msun$, we measure a constant value of $\fgas = 26\%$, so that galaxies with $\mstar > 10^{11}\,\msun$ actually have larger gas fractions than expected from the above trend. This can be explained if these galaxies also had lower $\sfe$s in the past, suggesting that we are witnessing a process that acts on long timescales.

We also find that the overall decrease of gas fraction cannot be explained solely from the growing mass of the bulges. Indeed, if we substitute the disk mass to the total stellar mass, using the average $\bt$ measured in each mass bin and assuming that galaxies of $\mstar < 10^{10}\,\msun$ are pure disks, the gas fraction in the disk is also found to decrease, albeit with a slightly shallower slope. Similar results are obtained if we use the $\bt$--$\mstar$ relations of \cite{lang2014}.

It should be noted that the $\sfe$ and $\fgas$ we measure in high-mass galaxies are consistent with the $z=1$ value reported by \cite{bethermin2015-a}, who applied the same methodology to a single mass bin around $\mstar \sim 10^{11}\,\msun$ using galaxies from the larger COSMOS field. On the other hand, similar measurements were performed in \cite{santini2014}, in the same field as \cite{bethermin2015-a}, finding smaller gas masses by about a factor of $3$. The discrepancy appears to come from different calibrations of the dust-to-gas ratio, and therefore should only result in a systematic shift. In any case, owing to the shallow depths of the COSMOS survey, \cite{santini2014} could only focus on galaxies more massive than $3\times10^{10}\,\msun$, i.e., they do not probe the linear Main Sequence regime (as is illustrated in \rfig{FIG:sfe}, right).

\begin{figure}
    \centering
    \includegraphics[width=9cm]{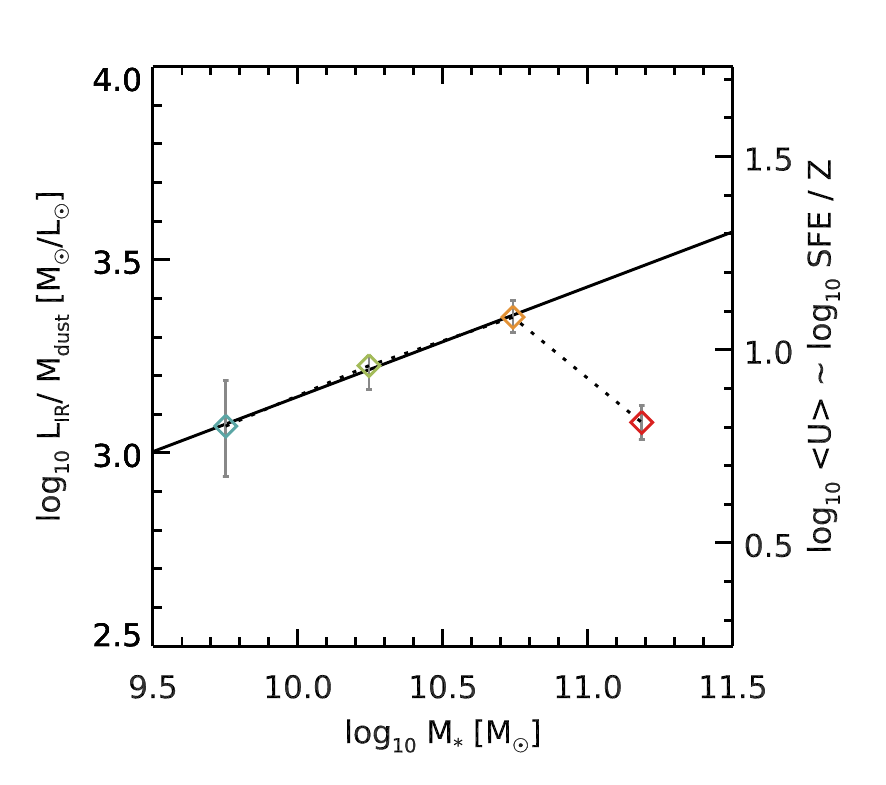}
    \caption{Ratio between the dust mass ($\mdust$) and the total infrared luminosity ($\lir$) as a function of the stellar mass for stacked galaxies at $z=1$. Colors are the same as in \rfig{FIG:sfe}. We overplot a linear fit (in log space) of the first three mass bins with a solid black line.}
    \label{FIG:lirmdust}
\end{figure}

Lastly, to see how the assumptions about metallicity and gas-to-dust ratio affect our result, we show in \rfig{FIG:lirmdust} the $\lir/\mdust$ ratio, which is a more direct observable. With our adopted dust model, neglecting the contribution of PAHs (n.b.: they represent only $4\%$ of the total dust mass), the following relation links together this ratio and $\tdust$, or equivalently, the mass-weighted average intensity of the stellar radiation field $\mean{U}$ to which dust grains are exposed:
\begin{equation}
\frac{\lir}{\mdust} \left[\frac{\lsun}{\msun}\right] = 185 \, \left(\frac{\tdust}{17.5\kelvin}\right)^{5.54} = 185 \, \mean{U}\,.
\end{equation}

The observed behavior of the $\lir/\mdust$ ratio is very similar to that of the $\sfe$, namely there is a steady rise with stellar mass, and then a sudden drop at $\mstar > 10^{11}\,\msun$. This should not come as a surprise, knowing that our estimated gas-to-dust ratio ends up being a simple power law of the stellar mass (see \rsec{TAB:stackprop}), and that the $\sfr$s in this sample are largely dominated by the dust-obscured, IR-luminous component. The low-mass slope that we find here is fairly shallow, although we rule out a flat slope (as reported in \citealt{magdis2012}) at the $3\sigma$ level. Yet, even if we were to adopt such a flat slope as the reference trend, the drop of $\lir/\mdust$ (or $\sfe$) in the highest mass bin would be less pronounced but still significant ($4\sigma$).

\subsection{A progressive and mass-dependent decrease of the $\sfe$ with time}

\begin{figure*}
    \centering
    \includegraphics[width=18cm]{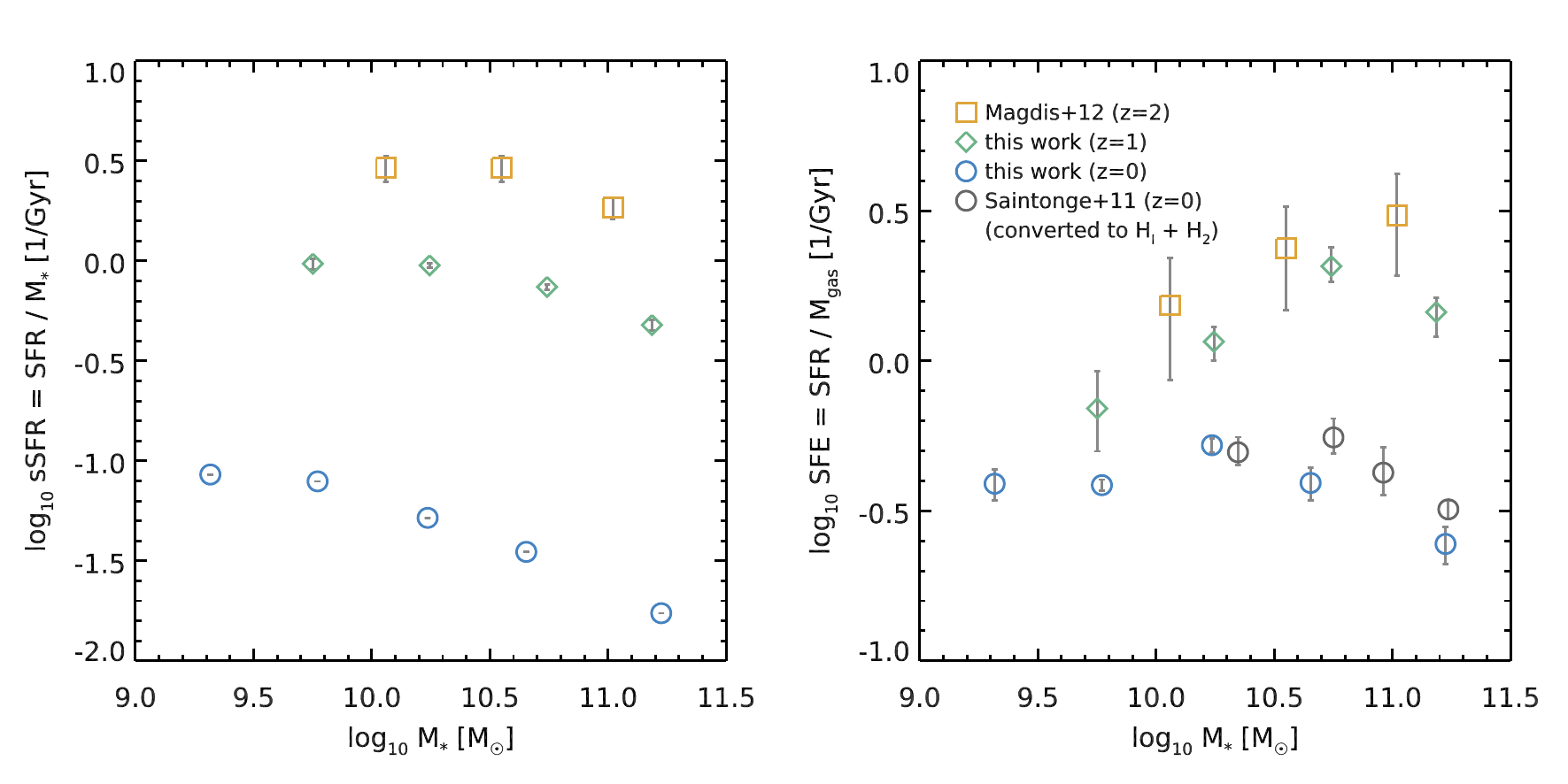}
    \caption{{\bf Left:}Relation between the specific $\sfr$ ($\ssfr=\sfr/\mstar$) and the stellar mass ($\mstar$), at various redshifts. Our $z=1$ stacked measurements from \citetalias{schreiber2015} are shown with empty diamonds, and the average values of the star-forming HRS galaxies are shown with empty circles. The associated error bar is the error on the mean, not the dispersion of the sample. We compare these measurements to the $z=2$ values obtained by \cite{magdis2012} for star-forming \bzk galaxies. {\bf Right:} Same as left, but replacing the $\ssfr$ by the star formation efficiency ($\sfe=\sfr/\mgas$). The diamonds and circles use the gas mass estimated in this paper, while the empty squares come from \cite{magdis2012}, and were computed with the same method. We also show for reference the measurements of \cite{saintonge2011-a} with empty gray circles. Their study only included $\htwo$ in $\mgas$, so we rescale their measurement to include $\hone$ assuming $R_{\rm mol} = \log_{10}(\mhtwo/\mhone) = 0.425\,(\log_{10}(\mstar/\msun) - 10.7) - 0.387$ (\citealt{saintonge2011}, Fig.~9 and Table~4). We also caution that the sample selection in \cite{saintonge2011-a} is different from ours.}
    \label{FIG:sfez}
\end{figure*}

In \rfig{FIG:sfez} (right) we put together our $\sfr$ and $\mgas$ measurements at both $z=1$ (previous section) and $z=0$ using galaxies from the HRS survey to display the evolution of the $\sfe$ with stellar mass and redshift. The values in the HRS are obtained by binning galaxies in stellar mass, and computing the mean $\sfe$ in each bin, since all the HRS star-forming galaxies are individually detected by \herschel, and therefore have individual gas masses estimates. These results are compared to that of \cite{magdis2012}, who performed a similar analysis in the GOODS fields, stacking galaxies in different bins of stellar mass from $\mstar = 10^{10}$ to $3\times10^{11}\,\msun$, but focusing on $z=2$ \bzk galaxies\footnote{They did stack galaxies at $z=1$, but did not separate them in different stellar mass bins. Also, since the \bzk selection only selects star-forming galaxies at $z=2$, they had to use another method to discard quiescent galaxies at $z=1$. To do so, they used a cut in \sersic index of $n < 1.5$ \citep[see e.g.,][and \rfig{FIG:sfrms_index}]{wuyts2011}. Because the associated selection effects are not obvious to determine, we prefer not to consider this data point in the present analysis, although the gas fraction they report is compatible with the one we measure here.}. The selection effects inherent to the \bzk classification are not very well understood, and it is known that this selection tends to affect the shape of the Main Sequence \citep{speagle2014}. With this caveat in mind, we proceed comparing these results to our data at $z=0$ and $z=1$.

The first thing to note is that the $\sfe$ at various redshifts is sytematically different, with higher redshift galaxies showing higher $\sfe$s. This fact is known, and will not be discussed any further \citep[see, e.g.,][]{genzel2010,combes2013,tacconi2013,santini2014,bethermin2015-a}.

Similarly to our $z=1$ sample, the most massive galaxies in the HRS ($\mstar > 10^{10}\,\msun$) are also found to have a reduced $\sfe$, thereby confirming the trend observed in the previous section. However, \cite{magdis2012} observe a fairly different picture than the one we present here, since their galaxies of all stellar mass are found to lie on the same $\sfr$--$\mgas$ relation, i.e., following a universal star formation law.

In fact, this is fully consistent with the observed evolution of the high-mass slope of the Main Sequence \citep[see, e.g., the comprehensive analysis of][]{gavazzi2015}, since at $z=2$ the $\sfr$--$\mstar$ relation is found to be almost linear (see \citetalias{schreiber2015} and \rfig{FIG:sfez}, left), indicating that whatever process drives this change of slope has not yet taken place. On the other hand, at $z=0$ the bending of the Main Sequence is more pronounced and takes place above a turnover mass that is lower than at $z=1$, in agreement with the behavior of the $\sfe$ that we observe for the HRS galaxies.

Similar trends of decreasing $\sfe$ with stellar mass have been reported in the literature \citep[e.g.,][]{saintonge2011-a,dessauges-zavadsky2015,mok2015}, although these studies do not mention a turnover of this relation. We argue that this is nevertheless consistent with our result, since these studies could only observe the regime \emph{above} the $z=0$ turnover mass, where the $\sfe$ is going down (see, e.g., \rfig{FIG:sfez} where we overplot the measurements of \citealt{saintonge2011-a}). Furthermore, it is also likely that this turnover of the $\sfe$--$\mstar$ relation can only be seen if the \emph{total} gas mass is used, i.e., including atomic hydrogen. Indeed, low-mass galaxies typically have lower $\mhtwo/\mhone$ ratios \citep[e.g.,][]{saintonge2011,boselli2014-a}, and would have substantially higher $\sfe$s if only molecular gas is used (see, e.g., \citealt{gardan2007,leroy2008,gratier2010,boselli2014-a}).

\section{Discussion}

\subsection{Quantifying the ``quenching'' and ``downfall'' rates}

\begin{figure}
    \centering
    \includegraphics[width=9cm]{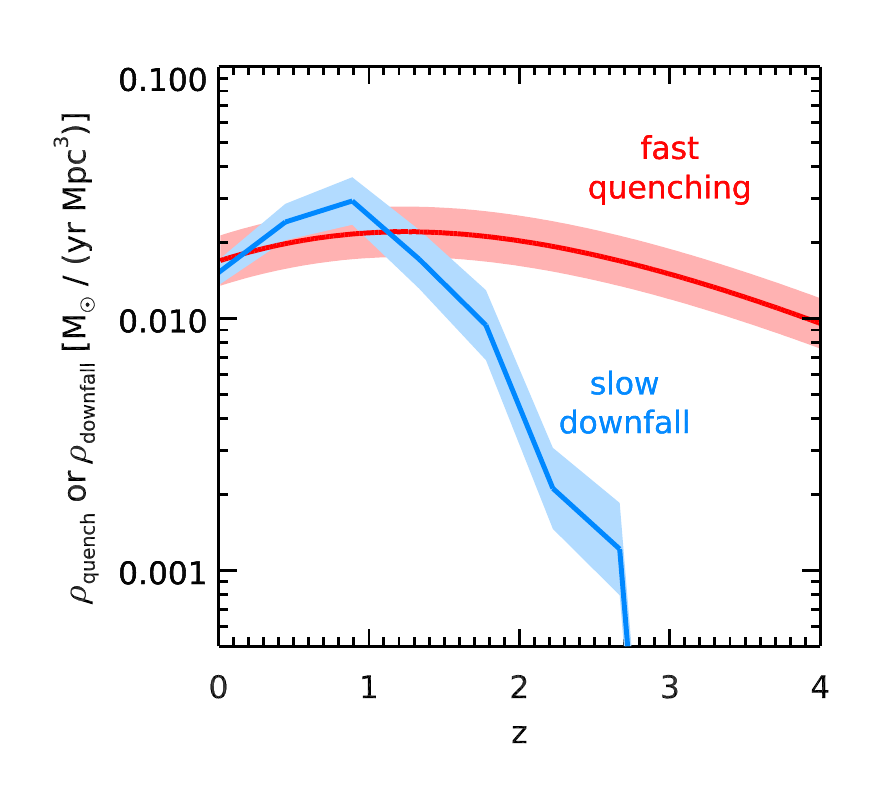}
    \caption{Evolution of the mass-weighted quenching and downfall rate densities with redshift. The red curve shows the time derivative of the stellar mass density of \uvj quiescent galaxies, which we assume are produced by a ``fast quenching'' mechanism. The blue curve shows the star formation density that is lost because of the lowered $\sfe$ in massive galaxies, which we call the ``slow downfall'' rate. The shaded regions in the background give the uncertainty on both measurements.}
    \label{FIG:rhoquench}
\end{figure}

We find that the bending of the Main Sequence cannot be caused by abnormally low gas fractions, but is instead resulting from a progressive decrease of the star formation efficiency, as shown in \rfigs{FIG:sfe} and \ref{FIG:sfez}. These observations converge toward a ``slow downfall'' of star formation, where massive galaxies gradually decrease their star formation activity while staying on the Main Sequence (see also \citealt{tacchella2015-a}). While staying on the Main Sequence, these galaxies become gradually less efficient in their star formation activity instead of abruptly turning off though a ``fast quenching''. Because the $\sfe$ is going down with time, these galaxies do not grow too massive by $z=0$, as shown in \cite{leja2015} who simulate the evolution of the observed stellar mass function using a Main Sequence of varying slope. The downfall of the star formation rate in massive Main Sequence galaxies may lead to the death of galaxies if, e.g., the gas surface density falls below the critical density that is necessary to switch on the Schmidt--Kennicutt relation, but our analysis does not allow us to make any firm claim favoring or disfavoring a scenario in which this downfall feeds the red sequence. Instead, we propose here to quantify the ``downfall rate'' of this slow process, and compare it to the fast quenching rate associated with the growth of the red sequence.

As shown, e.g., in \cite{muzzin2013} and \cite{tomczak2014}, the stellar mass density of \uvj quiescent galaxies increases monotonously with time, illustrating the progressive buildup of the red sequence. The time derivative of this quantity, neglecting stellar mass loss and residual star formation, is a measure of the quenching rate of galaxies \citep[see, e.g.,][]{peng2010}. Here, we make the hypothesis that all the \uvj quiescent galaxies were quenched by a \emph{fast} process, and set
\begin{align}
\rho_{\rm quench} &= \frac{\dd \rho_{*}^{\rm Q}}{\dd t}\,,
\end{align}
where $\rho_{*}^{\rm Q}$ is the stellar mass density of \uvj quiescent galaxies. We parametrize this latter quantity by fitting the redshift evolution reported in the CANDELS fields by \cite{tomczak2014}, accounting for the different choice of IMF:
\begin{align}
\rho_{*}^{\rm Q}\,\left[\msun/\Mpc^3\right] &= (2.6 \pm 0.7) \times 10^8 \exp(-z) \,.
\end{align}

To estimate the downfall rate associated to the \emph{slow} process that lowers the $\sfe$ of massive star-forming galaxies, we compute the difference between the \emph{observed} $\sfr$ density ($\rho_\sfr$) and the density that \emph{would be observed} if there was no drop of $\sfe$, therefore if the Main Sequence had a slope of unity at all stellar masses ($\rho_\sfr^{\rm unity}$). This is a measure of the amount of star formation that was lost because of the reduced $\sfe$ within the Main Sequence. We estimate both $\sfr$ densities using the stellar mass functions of star-forming galaxies introduced in \citetalias{schreiber2015} (that we complement toward $z=0$ using the mass function from \citealt{baldry2012}), and integrate these mass functions weighted by the $\sfr$. For the observed $\rho_\sfr$, we use the $\sfr$--$\mstar$ relation given in \citetalias{schreiber2015}. Defining $r \equiv \log_{10}(1+z)$ and $m \equiv \log_{10}(\mstar / 10^{9}\,\msun)$, this relation reads
\begin{eqnarray}
\log_{10}(\sfrms [\msun / {\rm yr}]) =  m - m_0 + a_0\,r \hspace{2.5cm} \nonumber \\
\hspace{0.9cm} - a_1 \, \big[{\rm max}(0, m - m_1 - a_2\,r)\big]^2\,,
\label{EQ:sfrms}
\end{eqnarray}
with $m_0 = 0.5 \pm 0.07$, $a_0 = 1.5 \pm 0.15$, $a_1 = 0.3 \pm 0.08$, $m_1 = 0.36 \pm 0.3$ and $a_2 = 2.5 \pm 0.6$. For $\rho_\sfr^{\rm unity}$ we use this same equation excluding the last term (which is used to describe the bending), i.e.:
\begin{eqnarray}
\log_{10}(\sfrms^{\rm unity} [\msun / {\rm yr}]) =  m - m_0 + a_0\,r\,.
\label{EQ:sfrms_unity}
\end{eqnarray}
Since these equations were not calibrated at $z<0.5$ in \citetalias{schreiber2015}, we use the observed Main Sequence from the HRS galaxies for these redshifts.

The downfall rate is then defined simply as
\begin{eqnarray}
\rho_{\rm downfall} = \rho_\sfr^{\rm unity} - \rho_\sfr\,.
\end{eqnarray}

The resulting evolution of both $\rho_{\rm quench}$ and $\rho_{\rm downfall}$ is shown in \rfig{FIG:rhoquench}. One can see from this figure that the fast quenching mode clearly dominates at all $z>1.5$, while the slow downfall rapidly catches up to reach similar rates from $z=1.5$ to the present day, i.e., over $\sim70\%$ of the history of the Universe.

Two conclusions can be drawn from this observation. First, the fact that both the quenching and downfall rates reach similar values at all $z<1.5$ implies that the downfall is a quantitatively important effect that should be considered alongside the growth of the red sequence. Second, it is clear that the two modes act at different epochs in the history of the Universe. While the fast quenching appears to hold a steady rate all the way from $z=4$ to the present day, the slow downfall becomes a significant source of SF suppression only at $z<2$. This suggests that the buildup of the red sequence and the change of slope of the Main Sequence are in fact related to two separate physical processes. This is discussed further in the next section.

\subsection{Identifying the actors that regulate the $\sfe$ and the gas content}

We show in \rsec{SEC:sfrmdisk} that the bending of the Main Sequence remains even if we are to consider only the stellar mass of the disk, excluding the inert bulges. While it is natural to expect that the specific star formation rate of galaxies could be universal only when computed over the disk rather than total mass of galaxies (as proposed by \citealt{abramson2014}) since bulges do not form stars, it would also generate a tension with another concept linked to the Main Sequence, namely the fact that galaxies are fed by the infall of extragalactic matter, which is in turn proportional to the total mass of galaxies including dark matter \citep[e.g.,][]{dekel2013}: the bulge, even if not forming star, does contribute to the gravitational potential of the galaxy, and must therefore provoke additional infall. Hence the fact that our results from \rsec{SEC:sfrmdisk} refute bulge growth as the actor of the Main Sequence bending may not be surprising, and possibly even expected when accounting for the large-scale context of infall. This also echoes the result obtained more recently in the SDSS by \cite{guo2015}, who also found a sublinear slope (i.e., less than unity) for the $\sfr$--$\mstar$ relation of $z=0$ pure disk galaxies, in conflict with the results of \cite{abramson2014}.

As discussed in the previous section, we observe instead in \rsec{SEC:sfemstar} that the star formation efficiency is decreasing in massive galaxies, leading to a slow downfall of star formation. This suggests the existence of an \emph{active} process that impacts the star formation activity, although the question remains to figure out exactly what this process could be. We cannot definitely address this question with the present data alone, but we review in the following the known mechanisms in light of our results.

We may already state that feedback from supernovae is not the favored solution, for it would affect more efficiently galaxies with a low gravitational potential, and therefore with low stellar masses, oppositely to our finding. Interestingly, the range in redshift and galaxy mass where the Main Sequence flattens corresponds to the regime where theory predicts group formation to be most effective, hence suggesting that structure formation or the membership to massive haloes may affect the rate of gas infall and the energetics regulating star formation (disk rotation and turbulence, see, e.g., \citealt{hennebelle2008}). Gravitational heating \citep{birnboim2003,dekel2008}, i.e., the injection of energy into the dark matter halo from gas accretion itself, only depends on the mass of this halo, and can therefore act also in isolated galaxies. According to \cite{dekel2008}, this can completely stop star formation in halos more massive than $\sim6\times10^{12}\,\msun$, corresponding to a typical stellar mass of $\sim2\times10^{11}\,\msun$ at $z=1$ \citep{behroozi2013}. This halo mass is the threshold above which natural cooling cannot counterbalance the energy brought into the halo by accretion, but in fact this energy is always there, even below this mass threshold, and can affect less massive halos more moderately (see, e.g., the scenario proposed by \citealt{tacchella2015} where galaxies with masses as low as $10^{10}\,\msun$ can be affected). Interestingly, it has been observed that AGN-driven outflows also act preferentially above a similar characteristic stellar mass: more than half of the star-forming galaxies above $\mstar > 10^{11}\,\msun$ show signs of such outflows, while this fraction drops below $20\%$ at $\mstar < 5\times10^{10}\,\msun$, at both $z=2$ and $z=1$ \citep{foersterschreiber2014,genzel2014}. While these winds have in principle enough energy to push the gas out of the galaxy, it is likely that they will also impact the distribution of the gas \emph{within} the galaxy, preventing fragmentation or disrupting molecular clouds. The reason why this would impact the $\sfe$ preferentially at $z\le1$ is unclear, although it could be linked to the fact that $z=2$ galaxies are more clumpy and gas-rich, and are therefore less affected by the winds \citep{roos2015}. On the other hand, we cannot rule out the action of the ``radio-mode'' AGN feedback, where jets heat the gas in the surroundings of galaxies, that may also be more common in massive galaxies.

Lastly, another key quantity that is related to the stellar mass is the metallicity. Indeed, it has been proposed that metallicity could be a main driver of the $\sfe$ at small scales, influencing the conversion of prestellar cores into stars through the strength of stellar winds, hence also setting the global $\sfe$ of the galaxy \citep[e.g.,][]{dib2011-a}. In their work, \cite{dib2011-a} predict a steady decrease of the molecular $\sfe=\sfr/\mhtwo$ with metallicity, which qualitatively match our observations at high stellar masses. At low stellar masses, the dominance of $\hone$ likely dilutes the effect predicted by Dib et al., which does not affect the conversion of $\hone$ into $\htwo$. Investigating this path further would require more precise metallicity measurement than what we used in the present paper.

Over the last years, the emphasis was put mostly on violent quenching mechanisms to explain the low baryonic fraction per unit dark matter halo mass, switching off the growth of galaxies by supernovae and AGNs at low and high masses, respectively (see, e.g., \citealt{silk2012,behroozi2013,behroozi2015}). We present here evidence that a slow downfall of the star formation efficiency should also be considered as a key mechanism.

\section{Conclusions}

We addressed here the origin of the change of slope of the Main Sequence of star-forming galaxies at $z<1.5$, where high-mass galaxies exhibit a lower $\ssfr\equiv\sfr/\mstar$ than what one would extrapolate from low-mass galaxies \citep[e.g.,][]{whitaker2012-a,magnelli2014,whitaker2014,ilbert2015,schreiber2015,lee2015,gavazzi2015}.

It was reported in the Local Universe that the $\sfr$--$\mdisk$ relation is linear, suggesting that it is the bulge that creates most of the change of slope of the Main Sequence \citep{abramson2014}. This claim was recently questioned by \cite{guo2015} at $z=0$, who reported that the slope of the $\sfr$--$\mdisk$ relation is in fact sublinear.

We performed the bulge-to-disk decomposition of a sample of $\sim1\,000$ galaxies at $z=1$ in the CANDELS fields, with robust $\sfr$s measured from their mid- to far-IR photometry. We find that, as for the $\sfr$--$\mstar$ relation, the high mass slope of the $\sfr$--$\mdisk$ relation remains substantially shallower than unity. Such shallow slope is also observed among pure disk galaxies, selected either from their decomposed bulge-to-total ratio, or from their effective \sersic index (see also \citealt{salmi2012} for a similar result at $z=1$). This implies the existence of a physical mechanism at play even within the disks of massive galaxies, uncorrelated to the presence or absence of a bulge.

We then used \herschel stacking to derive jointly the average $\sfr$ and dust mass of star-forming galaxies in four bins of stellar mass in the same redshift range.  Deriving the gas-phase metallicity from the Fundamental Metallicity Relation, we inferred the total gas mass, assuming that a fixed fraction of the metals are locked into dust, and analyzed the relation between the $\sfe\equiv\sfr/\mgas$ and the gas fraction in bins of stellar mass. We found that the most massive galaxies with $\mstar > 2\times10^{11}\,\msun$ show a significantly reduced $\sfe$ by about a factor of $2$ to $3$ when compared to extrapolations from lower stellar masses, while the gas fraction remains constant. We measured gas masses in Local galaxies from the \herschel Reference Survey and found a similar behavior, reinforcing this finding. There, the drop of $\sfe$ happens at lower stellar masses, in agreement with the redshift evolution of the slope of the Main Sequence (see \citetalias{schreiber2015}).

Combined together, these results point toward the existence of a \emph{slow downfall} mechanism that impacts the $\sfe$ of massive star-forming galaxies. We showed that this phenomenon is quantitatively important at $z<1.5$, and is likely disconnected from the fast quenching phenomenon that builds the red sequence. We argue that both mechanisms should be considered on the same footing when exploring the latest stages of galaxy evolution.

Leads for future research include studying the variation of the $\sfe$ above and below the Main Sequence, at fixed stellar mass. In this paper we show evidence that variations of $\sfr$ at high stellar masses are caused by variations of the $\sfe$ rather than gas mass. Since we have only been able to probe this through stacking and with relatively uncertain selection effects at $z=1$, it would certainly be interesting to confirm these trends for individual objects. This kind of analysis can only be accomplished using a statistically complete sample of $\sfr$ and dust mass measurements at different stellar masses (ideally with direct metallicity estimates from emission lines). While $\sfr$s and metallicities are currently within our reach, ALMA observations remain the only way to derive individual dust mass measurements for non-starbursting systems. A statistical sample with such measurement can be obtained either through dedicated pointed observations, or using a blind continuum survey, which will soon become possible with ALMA.

\begin{acknowledgements}
We thank the referee, J.~Braine, for his comments and suggestions that improved the readability, clarity and correctness of this paper.

CS wants to thank F.~Galliano for his input on the dust grain composition and for making available his dust model.

Most of the numerical analysis conducted in this work have been performed using {\tt phy++}, a free and open source C++ library for fast and robust numerical astrophysics (\hlink{cschreib.github.io/phypp/}).

This work is based on observations taken by the CANDELS Multi-Cycle Treasury Program with the NASA/ESA \hst, which is operated by the Association of Universities for Research in Astronomy, Inc., under NASA contract NAS5-26555.

This research was supported by the French Agence Nationale de la Recherche (ANR) project ANR-09-BLAN-0224 and by the European Commission through the FP7 SPACE project ASTRODEEP (Ref.No: 312725).

LC benefited from the {\sc thales} project 383549 that is jointly funded by the European Union and the Greek Government in the framework of the program ``Education and lifelong learning''.

T.W. acknowledges support for this work from the National Natural Science Foundation of China under grants 11303014.
\end{acknowledgements}

\bibliographystyle{aa}
\bibliography{../bbib/full}

\appendix

\section{Cleaning the $24\,\um$ catalogs \label{APP:24clean}}

We focus here on the association of a \spitzer MIPS $24\,\um$ flux to the galaxies in the $H$-band catalog. The procedure that was used to build the $24\,\um$ flux catalog \citep[see][]{magnelli2009} is based on IRAC $3.6\,\um$ position priors: sources are extracted on the $24\,\um$ map (and then, sequentially on the \herschel images) at the position of bright $3.6\,\um$ sources. If two priors are too close to be deblended on the MIPS image, only the brightest $3.6\,\um$ source is kept in the prior list. Because the IRAC bands are good tracers of the stellar mass, and because the stellar mass correlates with the star formation rate, this approach is very effective for extracting reliably the vast majority of the MIR and FIR sources. But it will fail in a few rare cases that will be particularly important for our study \citep[see also][]{mancini2015}. Indeed, one expects the method to be biased as soon as some objects deviate from the $\sfr$--$\mstar$ correlation. For example, it will happen that a massive, quiescent galaxy lies within a few arcseconds of a smaller mass (or slightly higher redshift) star-forming galaxy. The quiescent galaxy, being very massive, is most likely the brightest emitter in the IRAC $3.6\,\um$ image, however it is not expected to shine much in the MIR because it is not forming any stars. The nearby star-forming galaxy on the other hand can be fainter in the IRAC image, but will contribute to most, if not all, of the MIR emission. In this situation, the typical outcome is that the star-forming galaxy is removed from the prior list, since it has the faintest IRAC flux, while the quiescent galaxy is given all the IR flux. The end result is that we do have in our catalogs a few massive quiescent galaxies with bright $24\,\um$ emission that are obvious mismatches. We emphasize that the issue does not affect the $24\,\um$ fluxes listed in the published catalogs, but rather the association of these fluxes to counterparts in the higher-resolution \hst images.

We therefore eyeballed every galaxy of the $H$-sample that was attributed a counterpart in the MIPS image, looking for this kind of problematic cases. To identify quiescent galaxies, we rely on the \uvj classification introduced in the previous section. In total, we find $40$ clearly wrong associations over the four CANDELS fields, based on a combination of the \uvj classification and the presence of a likely star-forming candidate nearby, or by significant off-centering of the MIPS emission. Because this approach is hard to replicate and translate to other surveys, we introduce here a systematic and objective procedure to identify this kind of issues that does not require eyeballing every galaxy. It also allows us to further refine the flagging and discard not only galaxies that are clearly wrong associations, but also those that are uncertain, so that we work with a sample that is as clean as possible.

For each \uvj star-forming galaxy in the $H$-sample, we derive their expected ``Main Sequence'' star formation rate from their redshift and stellar mass, i.e., the $\sfr$ they would have if they were exactly following the Main Sequence as defined in \citetalias{schreiber2015}. From this $\sfr$ we subtract the observed, non-dust-corrected $\sfruv$, and use the \cite{kennicutt1998-a} relation to convert the remaining obscured $\sfr$ into $\lir$. We then use the best-fit IR SEDs of \citetalias{schreiber2015} to estimate their $24\,\um$ flux. For \uvj quiescent galaxies, we follow a similar procedure where the total $\sfr$ is instead taken from the stacking of \uvj quiescent galaxies, as described in the Appendix of \citetalias{schreiber2015}. This $\sfr$ is typically a factor of ten below the Main Sequence at all stellar masses\footnote{This may sound surprisingly high, but it should be noted that this stacked ``$\sfr$'' of quiescent galaxies also includes, for a large fraction, some $\lir$ coming from the dust headed by old stars, and not actual star formation. Therefore this prescription allows us to take into account both residual star formation and dust headed by old stars at the same time. See also \cite{fumagalli2014} where this was done in more details.}.

Using this procedure we are able to obtain a rough prediction of the MIR output of all the galaxies in the $H$-band parent sample. Then, for each galaxy with a $24\,\um$ detection, we estimate the reliability of the MIR association. To do so, we take all the galaxies that 1) lie within $4\arcsec$ of the detection, 2) have a \emph{predicted} $24\,\um$ flux that is at least a tenth of that predicted for the detection, and 3) have no measured $24\,\um$ (or below $3\sigma$) in the catalog. We then sum all their fluxes, weighted by the MIPS PSF amplitude at their corresponding distance, and divide this sum by the predicted flux of the detection. The resulting value gives an estimation of the fraction of the measured flux that can be contaminated by neighboring sources that were excluded from the prior list.

As expected, the vast majority of the sources in the MIPS catalog are classified as robust identifications: $80\%$ of them have an estimated contamination of zero. In this paper, we only use the individual $\sfr$s of galaxies for which this contamination fraction is below $30\%$. This criterion recovers $27$ of the $40$ wrong associations we identified by eye, the remaining $13$ galaxies are either not properly deblended on the \hst image, or their neighbors have wrong photometric redshifts and their contamination is underestimated. We therefore also exclude these $13$ galaxies from our sample.

Note that this flagging does \emph{not} apply to the sample we use to make the gas mass measurements (\rsec{SEC:sfe}). Indeed, the gas masses are measured by stacking $H$-band selected galaxies, and therefore do not rely on the $24\,\um$ catalogs.

\section{Robustness of the bulge-to-disk decomposition \label{APP:simu}}

\begin{figure}
    \centering
    \includegraphics[width=8cm]{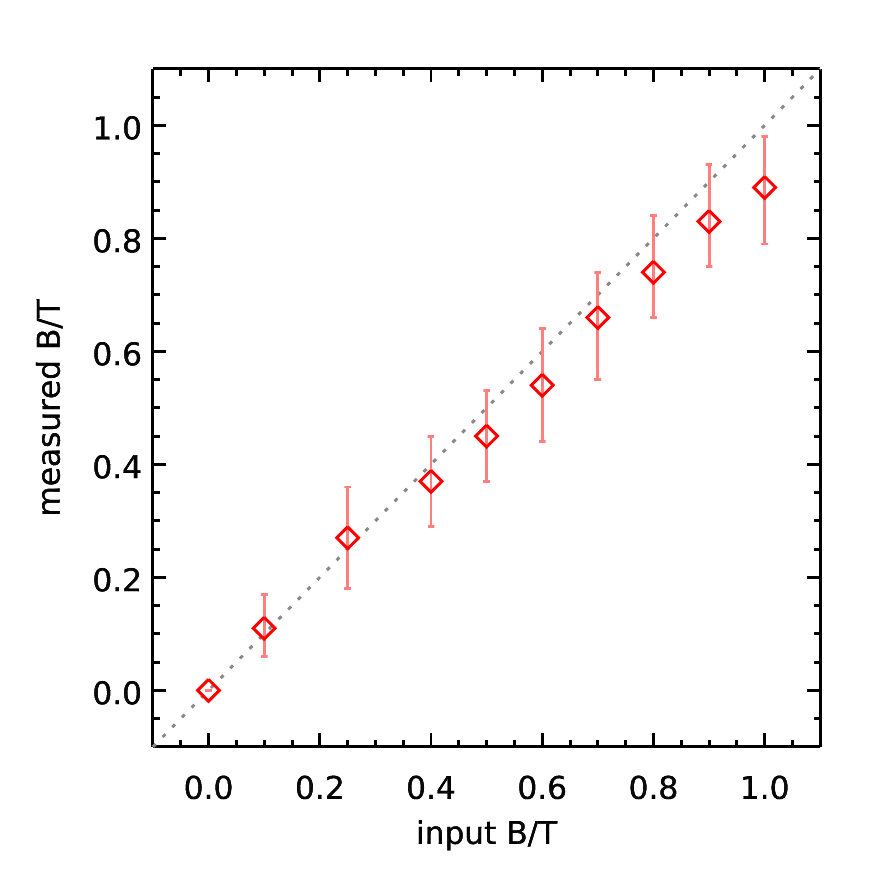}
    \caption{Comparison between the simulated $\bt$ and that measured by \gimtod, for galaxies with $H < 22.5$. The median measured $\bt$ are shown with empty red diamonds, and the error bars give the $16$th and $84$th percentiles of the distribution. The dotted line in the background gives the expected one-to-one relation.}
    \label{FIG:simu}
\end{figure}

To test the robustness and quality of our morphological decomposition, we create a large set of simulated galaxies of known profiles and $\bt$, and try to measure their properties in the presence of photometric noise. To do so, we use \galfit \citep{peng2002} to model $5\,000$ idealized double \sersic profiles ($n=1$ and $n=4$) of varying sizes, axis ratios, position angles, and fluxes, and place these models on empty regions of the real \hst images. We then run both \galfit and \gimtod trying to find back the input parameters.

We find that the total magnitude of the galaxy is always well recovered, except in the case of some catastrophic failures which happen almost exclusively with \galfit. Enforcing that the measured total magnitude is close to that chosen in input effectively gets rid of most of these poor fits. For the real galaxies, we choose to compare the measured total magnitude to that quoted in the CANDELS catalogs, and discard \galfit runs for which the difference is more than $0.5$ magnitudes.

We also find that the bulge-to-disk decomposition is usually hopeless at $H>23$, as the measured $\bt$ are either very noisy or systematically biased toward roughly equal partition of the flux. For galaxies brighter than $H=23$, we show in \rfig{FIG:simu} the comparison between the $\bt$ we put in the simulation, and the ones that are recovered by \gimtod. We find that the code is able to identify disk dominated galaxies with great accuracy, while bulge-dominated galaxies and intermediate systems show a slight systematic underestimation: given the choice, \gimtod will tend to put more flux in the disk component than in the bulge. This effect is small however, and we checked that our conclusions are not affected if we correct for it by adding $0.05$ to the $\bt > 0.5$. We also observe that the uncertainty on the flux of the disk depends on $\bt$, with brighter bulges leading to more uncertain disk fluxes. For example, assuming constant mass-to-light ratio, for $\mdisk \simeq 2\times10^{10}\,\msun$, the error on $\mdisk$ is $0.04\,\dex$ for $\bt \simeq 0$, and $0.07\,\dex$ for $\bt > 0.3$. It should be noted that these simulations are only able to capture the ability of the codes to recover what was put on the simulated image, i.e., idealized profiles with realistic photometric noise and neighbor contamination, but it does not allow us to say how reliable is the decomposition in the case of perturbed, irregular or clumpy galaxies, nor does it hint about actually measuring a disk \emph{mass} (which is done in \rsec{SEC:mdisk}), e.g., it does not contain varying mass to light ratios. Therefore the real uncertainties on the measurements are probably larger. Still, even doubled, the errors we estimate here are low enough for our purposes.

The issue of this simulation approach is that we can only test our procedure against idealized galaxy profiles. To make sure that our results are not strongly biased by our decomposition approach, we also run in parallel the same decomposition of the real, observed profiles using \galfit. The same images and segmentations are used, the only difference is that we can allow for some small position offset between the bulge and the disk. The minimization procedure is also different between both codes, and therefore different results are usually obtained for the same data, providing an estimation of the uncertainty on the decomposition. Since \galfit requires an initial guess of the fit parameters, we used the single-component morphological parameters measured by \cite{vanderwel2012} who fit a single \sersic profile to the $H$-band image of each galaxy in the CANDELS catalogs of GOODS--{\it South}, UDS and COSMOS. We complement these measurements by running ourselves similar fits in GOODS--{\it North}. These parameters are used to set the initial size, axis ratio and position angle of both the disk and the bulge components, while the initial flux of each component is set to half the total flux of the galaxy (i.e., an initial $\bt=0.5$). We then run \galfit, leaving free every parameter including the position of each component, with a maximum offset between both components of $10$ pixels (in practice, the results are essentially the same if we do not allow for such offsets).

We have checked that our conclusions are not affected if we only keep the galaxies for which the two codes agree (variation of $\bt$ smaller than $0.15$), or if we used only the decomposition provided by \galfit. In the end, we prefer to used the results provided by \gimtod since this code does not require choosing starting conditions, which are known to influence strongly the final result of \galfit owing to the presence of local minima in the $\chi^2$ \citep[e.g.,][]{lang2014}. We also compared our results against the values obtained by running \mbox{Mega}\mbox{Morph} (\citealt{haeussler2013}; B.~H\"aussler, private communication). Since MegaMorph does not force the \sersic index of the bulge component to be equal to $n_{\rm bulge}=4$, we only perform the comparison against galaxies that MegaMorph chose to fit with $n_{\rm bulge} > 2$. We find a scatter in $\bt$ of about $20\%$, consistent with that found when comparing the results of \galfit and \gimtod.

\section{Impact of the \uvj selection on the gas mass measurements \label{SEC:uvj}}

\begin{figure*}
    \centering
    \includegraphics[width=9cm]{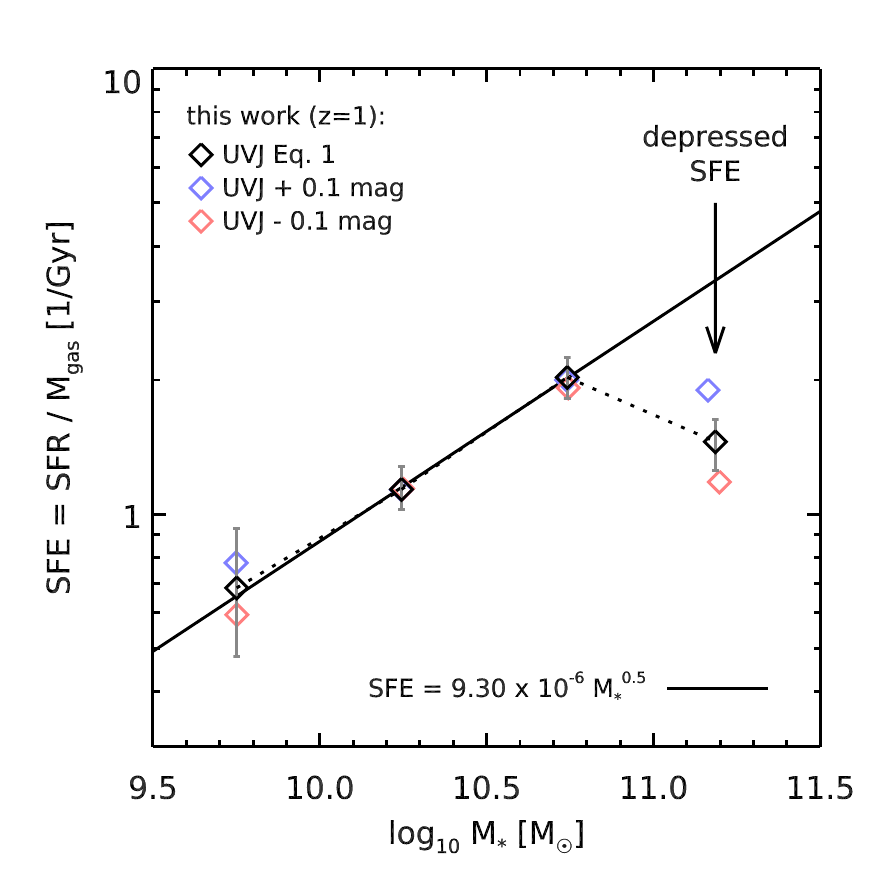}
    \includegraphics[width=9cm]{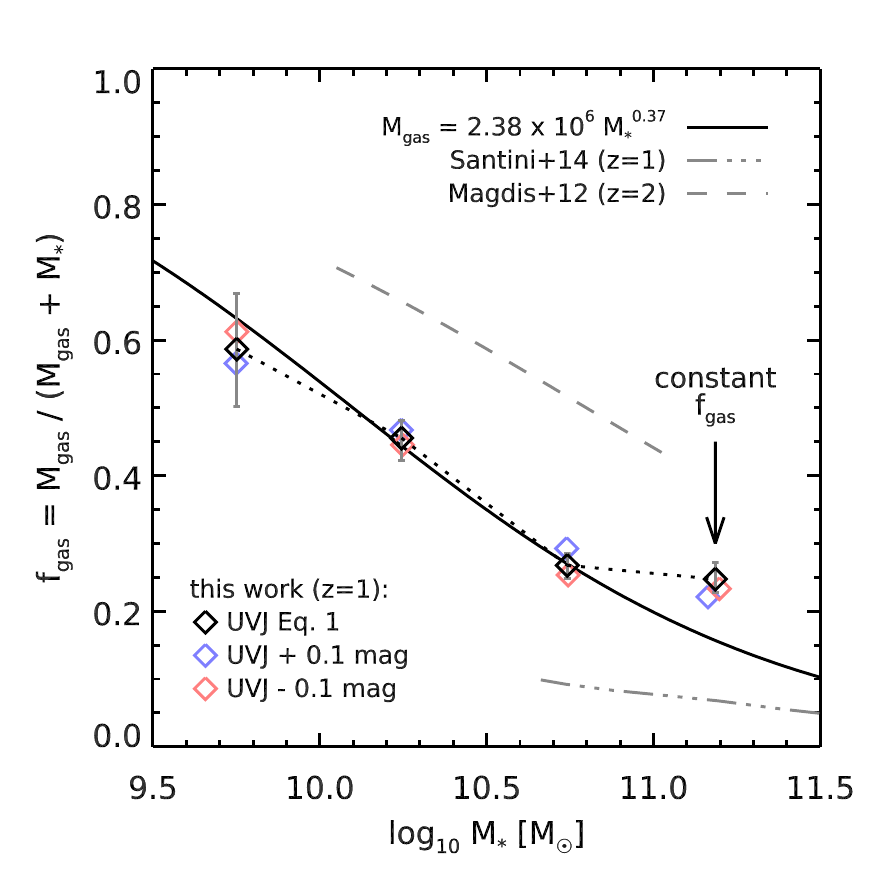}
    \caption{Same as \rfig{FIG:sfe}, but here black diamonds show the measured $\sfr$s and $\mgas$ of our chosen sample, while blue (respectively red) diamonds show how these values change if we shift the \uvj dividing line toward the star-forming (respectively quiescent) region by $0.1$ magnitude.}
    \label{FIG:sfe_uvj}
\end{figure*}

It has been shown that the properties of the $\sfr$--$\mstar$ relation, i.e., its slope but also its scatter, are very sensitive to the sample selection \citep[e.g.,][]{speagle2014}. In the present paper, we have used the standard \uvj color-color diagram to isolate quiescent galaxies, and although this selection has been widely used in the recent literature \citep[e.g.,][]{whitaker2012-a,muzzin2013,bruce2014,whitaker2014,lang2014,straatman2014,pannella2015}, its reliability can still be questioned. Indeed, while the quiescent and star-forming clouds can be easily identified on this diagram (see, e.g., \rfig{FIG:uvj_bt}), there is a non-negligible amount of galaxies in between, populating what is often referred to as the ``green valley''. The dividing line defined by \cite{williams2009} goes arbitrarily through this population, and it would be unwise to consider blindly that a ``green valley'' galaxy slightly above that line is quiescent, and that a similar galaxy slightly below the line is star-forming.

One way to circumvent this issue is not to apply any selection of star-forming galaxies in the first place, and identify the Main Sequence as the ridge (or mode) of the distribution of galaxies on the $\sfr$--$\mstar$ plane. This was done, e.g., in \cite{magnelli2014} and \cite{renzini2015}. However this approach is only feasible in samples that are not $\sfr$-selected. Building such a sample requires using $\sfr$s that are not fully based on the FIR, and that are therefore potentially unreliable \citep[one exception is the deep H$_\alpha$ data of the SDSS, as in][but translating this study to the distant Universe is currently out of our reach]{renzini2015}. Of course, this is also not applicable to stacking analyses, for which the $\sfr$ is only determined \emph{a posteriori}.

Coming back to the \uvj selection, there are two ways our study could be affected by this arbitrary dividing line. On the one hand, the selection may be too strict, and we could actually discard from our sample some galaxies that are still forming stars at non-negligible rates, but have colors similar to that of quiescent galaxies because of peculiar combination of star formation history and dust content. On the other hand, the selection may be too loose, and our ``star-forming'' sample could actually contain a number of quiescent galaxies. We expect both effects to take place mostly for the most massive galaxies, where dust is more abundant and where most quiescent galaxies are found. The first alternative can be addressed by looking at the position of \uvj quiescent galaxies in the $\sfr$--$\mstar$ plane. We see that there are indeed a few genuinely star-forming galaxies that are classified as \uvj quiescent. However, these galaxies tend to have systematically lower star formation rates compared to \uvj star-forming galaxies. Therefore, including these mistakenly identified galaxies in our sample would likely flatten the Main Sequence even more. The second alternative is probably more worrisome, as the drop of the $\sfe$ we observe in massive galaxies could be created by quiescent galaxies polluting our sample. One interesting observation to make out of \rfig{FIG:sfrms_bt} \citep[and that can be made more quantitatively by studying the distribution of $\sfr$ around the median value,][]{ilbert2015,schreiber2015} is that the mode of the $\sfr$ distribution at a given stellar mass (approximated here by the running median) coincides with the average value obtained from the stacked measurements. This means that, although our sample is $\sfr$-selected, the amount of galaxies below our $\sfr$ detection limit is small enough that their impact on the average trend is marginal. In fact, for galaxies more massive than $5\times10^{10}\,\msun$, where the bending of the sequence is most pronounced, $79\%$ of the \uvj star-forming galaxies are detected in the FIR. Therefore, the contamination of genuinely quiescent galaxies to the \uvj star-forming sample in this stellar mass range must be reasonably small (i.e., a maximum of $20\%$).

Nevertheless, in an attempt to quantify how our results are influenced by the choice of the \uvj dividing line, we replicate our $\sfe$ measurements by stacking two different additional samples which are built by slightly shifting the \uvj dividing line by $\pm 0.1$ magnitude. The resulting $\sfe$ and $\fgas$ are shown in \rfig{FIG:sfe_uvj}. As can be seen from this figure, moving the dividing line further into the quiescent cloud (red points) or further into the star-forming cloud (blue points) does not impact $\fgas$ in any statistically significant way. In both cases, we still observe a drop of $\sfe$, although the amplitude of this drop does vary, in this case mostly because of a change of $\sfr$.

This can be put in perspective with the work of \cite{arnouts2013}, who found that the $\ssfr$ of a galaxy could be inferred from its position on the $NrK$ diagram, which is conceptually similar to the \uvj diagram\footnote{By using rest-frame wavelengths that are further apart, this diagram has a larger dynamic range and will separate quiescent and star-forming galaxies more clearly than the \uvj diagram. The downside is that measuring the rest-frame $K$ band is particularly difficult at high redshifts, while the near-UV is hardly accessible at low redshift.}, with an $\ssfr$ that is continuously increasing as a function of the distance to the dividing line. According to \cite{arnouts2013}, using a stricter \uvj selection should bias our sample toward galaxies with a higher $\ssfr$, hence, at fixed mass, with a higher $\sfr$, which is what we observe for the most massive bin. In this context, the fact that the gas mass does not change substantially is particularly interesting, and is another hint that the mechanism responsible for the downfall, whatever it is, is mostly impacting the $\sfe$, and not the gas supply.

\end{document}